\documentclass[prd,twocolumn,nofootinbib,amssymb,floatfix,preprintnumbers,superscriptaddress]{revtex4}
\usepackage{graphicx,amsmath,mathtools}
\usepackage{epsfig,subfigure}
\usepackage{dcolumn}
\usepackage{bm}
\usepackage{float}
\usepackage{multirow}
\usepackage{cancel}

\def\beq{\begin{equation}}
\def\eeq{\end{equation}}
\def\bea{\begin{eqnarray}}
\def\eea{\end{eqnarray}}

\def\3Eqs#1#2#3{Eqs.\ (\ref{#1}), (\ref{#2}) and (\ref{#3})}

\def\to{\longrightarrow}

\def\within#1to#2/{#1\mbox{ to }#2}
\def\lbullet#1,#2,#3,#4/{\Text(#1,#2)[c]{$\bullet$}\Line(#1,#2)(#3,#4)}
\def\textcite#1{Ref.~\cite{#1}}
\def\sL#1{{\tilde L}_{#1}}
\def\sE#1{{\tilde E}_{#1}}
\def\newu{$U(1)_{L_\mu-L_\tau}$}
\def\newa{$U(1)_{L_\mu+L_\tau}$}

\def\gino{\tilde{g}}
\def\bino{\tilde{B}}
\def\bpino{\tilde{B^{\prime}}}
\def\wino{\tilde{W}}
\def\wpino{\tilde{W^{+}}}
\def\wmino{\tilde{W^{-}}}

\def\hd0ino{\tilde{h^0_{d}}}
\def\hdmino{\tilde{h^-_{d}}}

\def\hupino{\tilde{h^+_{u}}}
\def\hu0ino{\tilde{h^0_{u}}}
\def\etaino{\tilde{\eta}}
\def\etabarino{\tilde{\bar{\eta}}}
\def\sU{{\tilde U}}
\def\sD{{\tilde D}}
\def\sQ{\tilde Q}
\def\sEm{{\tilde E}_{\mu}}
\def\sEt{{\tilde E}_{\tau}}
\def\sEe{{\tilde E}_e}
\def\sLm{{\tilde L}_\mu}
\def\sLt{{\tilde L}_\tau}
\def\sLe{{\tilde L}_e}
\def\sneue{{\tilde \nu}_{e}}
\def\sneueR{{\tilde \nu}_{e R}}
\def\sneueI{{\tilde \nu}_{e I}}
\def\sneumu{{\tilde \nu}_{\mu}}
\def\sneumuR{{\tilde \nu}_{\mu R}}
\def\sneumuI{{\tilde \nu}_{\mu I}}
\def\sneutau{{\tilde \nu}_{\tau}}
\def\sneutauR{{\tilde \nu}_{\tau R}}
\def\sneutauI{{\tilde \nu}_{\tau I}}
\def\se{{\tilde e}_L}
\def\smu{{\tilde \mu}_L}
\def\stau{{\tilde \tau}_L}
\def\seR{{\tilde e}_R}
\def\smuR{{\tilde \mu}_R}
\def\stauR{{\tilde \tau}_R}
\def\etabar{\bar \eta}
\def\e2spl{\ensuremath{^\clubsuit}}
\def\diag{\mathop{\rm diag}}

\def\mET{E_T \hspace{-1.2em}/\;\:}

\def\nn{\nonumber\\*}

\begin{document}
\title{Supersymmetric gauged \boldmath U(1)$_{L_{\mu}-L_{\tau}}$ model for neutrinos and the muon (g-2) anomaly}

\author{Heerak Banerjee}
\email{tphb@iacs.res.in}
\affiliation{School of Physical Sciences, Indian Association for the Cultivation of Science,
2A $\&$ 2B Raja S.C. Mullick Road, Kolkata 700 032, India}
\author{Pritibhajan Byakti}
\email{priti137@gmail.com}
\affiliation{Department of Physics, 
Pandit Deendayal Upadhyaya Adarsha Mahavidyalaya (PDUAM) Eraligool, Karimganj-788723, Assam, India}
\author{Sourov Roy}
\email{tpsr@iacs.res.in}
\affiliation{School of Physical Sciences, Indian Association for the Cultivation of Science,
2A $\&$ 2B Raja S.C. Mullick Road, Kolkata 700 032, India}

\date{\today}

\begin{abstract}
The gauged \newu model can provide for additional contributions to the muon anomalous magnetic moment by means of a loop involving the $Z^{\prime}$ gauge boson. However, the 
parameter space of such models is severely constrained if one combines the latest muon $(g-2)$ data with various neutrino experiments, such as neutrino trident production, 
$\nu -e$ and $\nu -q$ elastic scattering, etc. In a supersymmetric \newu model, a larger region of parameter space opens up, thus enabling one to explore otherwise 
forbidden regions of parameter space in nonsupersymmetric models involving the new gauge coupling ($g_X$) and the mass of the $Z^\prime$ gauge boson ($M_{Z^{\prime}}$) . We show that the minimal 
model with the minimal supersymmetric Standard Model (MSSM) field content is strongly disfavored from $Z$-boson decay and neutrino data. We also show that the nonminimal model with two extra singlet superfields 
can lead to correct neutrino masses and mixing involving both tree-level and one-loop contributions. We find that, in this model, both muon $(g-2)$ and neutrino data may 
be simultaneously explained in a parameter region consistent with experimental observations. In addition, we observe that the muon $(g-2)$ anomaly can be accommodated even with higher 
values of electroweak sparticle masses compared to the MSSM. Charged lepton-flavor-violating processes (like $\mu\rightarrow e\gamma$, $\tau\rightarrow \mu\gamma$, etc.) may have 
potentially large branching ratios in this scenario. Depending on the magnitude of the supersymmetry contribution to these processes, they may constrain hitherto unconstrained 
regions of the $M_{Z^{\prime}}-g_X$ parameter space. However, we find that these branching fractions never exceed their upper bounds in a region where both muon $(g-2)$ and 
neutrino oscillation data can be simultaneously accommodated. 
\end{abstract}

\maketitle

\section{Introduction}

The Standard Model (SM) of particle physics is a successful theory. However, it does not seem to be a complete one: it cannot explain either the neutrino masses and mixing pattern or the $3.6 \sigma$ discrepancy between the SM prediction of the anomalous magnetic moment of the muon and its experimental value \cite{Bennett:2004pv, Bennett:2006fi, Miller:2007kk, Olive:2016xmw}. In order to explain the neutrino mass pattern and mixing \cite{Esteban:2016qun} and the muon anomalous magnetic moment, among other issues, one needs to look for physics beyond the SM. There are two basic ways in which the SM may be extended to address these issues: extend the field content of the SM, or extend the SM gauge symmetries. The simplest way to implement the latter is to enlarge the SM gauge group with an extra $U(1)$ gauge symmetry.

Out of several models available in the literature, a very interesting $U(1)$ extension that has attracted a lot of attention recently is the gauged \newu extension of the SM. It was first studied in the three-generation minimal Standard Model of quarks and leptons in the absence of right-handed neutrinos\cite{He:1990pn,He:1991qd}. The contribution of the extra gauge boson $Z^\prime$ of this model to the muon anomalous magnetic dipole moment was studied in Ref.\cite{Baek:2001kca}. The neutrino mass pattern and mixing angles in this class of models, with suitable field content, were discussed in Ref.\cite{Ma:2001md}, where the authors also discussed signatures of this model in high-energy colliders alongside an analysis of the muon $(g-2)$ anomaly.

A detailed fit to electroweak data was performed in Ref.\cite{Heeck:2011wj} in order 
to identify the allowed ranges of the mass of $Z^\prime$ and its mixing with the SM $Z$ 
boson. The authors also studied this model in the context of neutrino mass model building. 
Constraints on the mass and the coupling of the new gauge boson were derived from neutrino 
trident production in \cite{Altmannshofer:2014pba}. Dark matter candidates in this class of 
models and associated physics were discussed in Ref.\cite{Baek:2008nz,Baek:2015fea,Altmannshofer:2016jzy,Biswas:2016yan,Biswas:2016yjr,Patra:2016shz,Arcadi:2018tly,Biswas:2017ait}.
The possibility of detecting the gauge boson (assuming its mass in the range MeV$-$GeV) of \newu 
symmetry at the Belle-II experiment was discussed in Refs.\cite{Araki:2017wyg,Kaneta:2016uyt}. 
In addition, constraints on such a light gauge boson have been imposed from neutrino beam 
experiments \cite{Kaneta:2016uyt}, lepton-flavor-violating $\tau$ decays \cite{Chen:2017cic}, and rare kaon decays \cite{Ibe:2016dir}. 
Higgs boson flavor-violating decays were studied in Refs.\cite{Crivellin:2015mga,Altmannshofer:2016oaq}. 
Some recent anomalies involving $B$-meson decays were addressed in Refs.\cite{Altmannshofer:2014cfa,Crivellin:2015mga,Altmannshofer:2016jzy,Ko:2017yrd,Baek:2017sew,Arcadi:2018tly}. Neutrino masses and mixing were studied in a \newu$-$symmetric model with additional scalars and vector-like leptons in Ref.\cite{Chen:2017gvf} and with right-handed neutrinos in Ref.\cite{Baek:2015mna}. 

Considering the constraints on the gauged \newu model (especially from neutrino experiments), it was shown, for example, in Refs.\cite{Altmannshofer:2014pba,Araki:2017wyg}, that the parameter space allowed by the muon $(g-2)$ anomaly is severely restricted. Supersymmetry (SUSY) can be of immense help under these circumstances. The supersymmetric contribution to muon $(g-2)$ in conjunction with the usual contribution from $Z^{\prime}$ loop allows us to explore parameter spaces where any one of these contributions might be too low but together can explain experimental data quite well. Along these lines, we show that, in a SUSY version of the \newu model, a much larger region of the parameter space is free from all the experimental constraints, including those from the LHC, while still maintaining all of the attractive features of the non-SUSY version. The allowed parameter space in this scenario, which is otherwise forbidden in non-SUSY models, can be probed using various pertinent observables like neutrino masses and mixing, dark matter,  several $B-$decay anomalies and signatures at the LHC.

The minimal supersymmetric Standard Model (MSSM) is one of the most famous extensions of the SM \cite{Fayet:1976et, Martin:1997ns}. Although it introduces contributions 
unique to supersymmetry in muon $(g-2)$, data from the latest LHC experiments restrict the sparticle masses to higher and higher values\cite{Tanabashi:2018}. This makes 
explaining muon $(g-2)$ in MSSM increasingly difficult\cite{Hagiwara:2017lse}. In addition, $R-$parity-conserving MSSM cannot explain the tiny nonzero masses of the neutrinos 
and their nontrivial mixing pattern as observed in experiments involving solar, atmospheric, accelerator and reactor neutrinos. An intrinsically supersymmetric way of 
generating the small neutrino mass pattern and mixing is to introduce $R-$parity violation (RPV) (For a review, see, for example, Ref.\cite{Barbier:2004ez}). Another way of extending 
the MSSM, to accommodate neutrino masses and mixing is to enlarge the gauge group structure, and the simplest possibility is to augment the SM gauge group with an additional 
$U(1)$ symmetry. 

As we shall show, in a supersymmetric gauged \newu model it is possible to have much higher neutralino and slepton masses while still explaining muon $(g-2)$, in contrast with the MSSM as discussed above. In this work, we show how nonzero neutrino masses and a nontrivial mixing pattern can be achieved in this kind of a setup along with a prediction for the muon $(g-2)$ anomaly consistent with experimental observations. Let us note in passing that a supersymmetric version of \newu was also studied earlier in Ref.\cite{Das:2013jca} where the authors focused mainly on obtaining a leptophilic dark matter candidate in order to explain the PAMELA\cite{Adriani:2008zr} and AMS-02\cite{Aguilar:2013qda} results. 

Given that the explanation of neutrino 
mixing is directly connected to the presence of lepton flavor violation, we look into the processes $l_j\rightarrow l_i\gamma$ in particular, as these follow from a similar 
set of diagrams that give rise to $(g-2)_{\mu}$. We calculate the branching ratios of these processes in the nonminimal model and show that---because they are connected directly to both neutrino masses and the muon $(g-2)$ anomaly---they can never be too large where neutrino masses can be small enough while still explaining $(g-2)_{\mu}$. 
 
The plan of the paper is as follows. In Sec. \ref{sec:minimal} we describe the minimal model and discuss its essential features. The limitations of the minimal model will also be presented. The nonminimal model will be introduced in Sec. \ref{sec:nonminimal} and the scalar sector 
of the model will be studied. Sec. \ref{sec:neutrino} will be devoted to the fermionic sector of this model and the neutralino mass
matrix will be presented. We show how mixing of the neutrinos with the neutralinos along with the mixing involving the singlet fermions 
can generate tiny masses for the neutrinos. The mixing of the light neutrinos will be studied in detail in this section. The muon anomalous magnetic moment $(g-2)_\mu$ 
will be studied in Sec. \ref{sec:madm}. A detailed numerical analysis and allowed regions of the parameter space will be presented. A brief outline of the calculation 
of branching ratios for lepton-flavor-violating processes follow in Sec. \ref{sec:lfv}, along with a discussion of the results. Our conclusions and future directions 
will be presented in Sec. \ref{sec:conc}. Analytical expressions for the scalar mass matrices, the chargino mass matrix and the amplitudes for charged lepton-flavor-violating processes in the nonminimal model are included in the appendices.

\section{The minimal model}\label{sec:minimal}
The chiral superfield content of the minimal model is that of the MSSM with the gauge group $SU(3)_c \times SU(2)_L \times U(1)_Y \times U(1)_{L_\mu - L_\tau}$. 
The $U(1)_{L_\mu - L_\tau}$ charge assignments of different chiral superfields are shown in Table \ref{table-min-model-charges}. 
\begin{table}[htb]
\centering
\begin{tabular}{|c|ccccccccccc|}
\hline\hline
Superfields & ${\hat Q}_i$ & ${\hat U}^c_i$ & ${\hat D}^c_i$ & ${\hat L}_e$ & ${\hat E}^c_e$ & ${\hat L}_\mu$ & ${\hat E}^c_\mu$ & ${\hat L}_\tau$ & ${\hat E}^c_\tau$ & 
${\hat H}_u$ & ${\hat H}_d$ \\
\hline
$U(1)_{L_\mu - L_\tau}$ & 0 & 0 & 0 & 0 & 0 & 1 & -1 & -1 & 1 & 0 & 0 \\
\hline
\end{tabular}
\caption{$U(1)_{L_\mu - L_\tau}$ charge ($Q_X$) assignments to the chiral superfields of the minimal model}
\label{table-min-model-charges}
\end{table}
With the given charge assignments, 
we can write the following superpotential.
\begin{eqnarray}\label{e:mini:superpot}
&&W^{min}=\nn &&\epsilon_{ij}\bigg[-y^u_{nm} {\hat H}_u^i {\hat Q}_n^j {\hat U}_m^c + y^d_{nm} {\hat H}_d^i {\hat Q}_n^j {\hat D}^c_m + y^e {\hat H}_d^i {\hat L}_e^j {\hat E}_e^c \nn &&+ y^\mu {\hat H}_d^i {\hat L}_\mu^j {\hat E}^c_\mu + y^\tau {\hat H}_d^i {\hat L}_\tau^j {\hat E}^c_\tau - \mu_e {\hat L}_e^i {\hat H}_u^j - \mu_0 {\hat H}_d^i {\hat H}_u^j \nn 
&&+ \lambda_{122} {\hat L}_e^i {\hat L}_\mu^j {\hat E}^c_\mu + \lambda_{133} {\hat L}_e^i {\hat L}_\tau^j {\hat E}^c_\tau + \lambda'_{1nm} {\hat L}_e^i {\hat Q}_n^j {\hat D}^c_m \bigg].
\end{eqnarray}
Here the lepton flavor indices are explicitly written for each individual flavor. We have considered baryon number parity so that $\lambda^{\prime \prime}_{ijk} U^c_i D^c_j 
D^c_k$ is not allowed. The presence of \newu symmetry makes the Yukawa matrix for the lepton sector flavor 
diagonal.
The gauge symmetries alone dictate the pattern or nonzero elements of the 
couplings for lepton-number-violating terms as follows:
\begin{eqnarray}
\lambda_{212},\lambda_{122}, \lambda_{313}\mbox{ and } \lambda_{133} &\ne&0 \\
\lambda^{'}_{1jk} &\ne& 0\quad \forall\, j \& k \\
\mu_e &\ne& 0.
\end{eqnarray} 

The above superpotential has an accidental global 
symmetry: \newa. The soft SUSY-breaking terms for this model are as follows:
\begin{widetext}
\begin{eqnarray}\label{e:mini:soft}
-\mathcal{L}_{soft}^{min}&=&\frac{1}{2}\left(M_3(i\gino)(i\gino) +M_2(i\wino)(i\wino) +M_1 
(i\bino)(i\bino) +M_0 (i\bpino)(i\bpino) + h.c\right)- M_{10}(i\bino)(i\bpino) \nn
&&+\bigg(A_u^{ij} H_u \sQ_j\sU_i - A_d^{ij} H_d \sQ_j\sD_i  - A_e H_d\sLe\sEe  - A_{\mu} H_d\sLm\sEm -A_{\tau} H_d\sLt\sEt \nn
&&- A_{\lambda^{\prime}}^{eij} \sLe\sQ_i\sD_j - A_{\lambda_{122}} \sLe\sLm\sEm - A_{\lambda_{133}} \sLe\sLt\sEt  + h.c\bigg)+ M^2_{\sQ}{\sQ}^{\dagger}\sQ + M^2_{\sU} {\sU}^{c\dagger}\sU^c \nn
&&+ M^2_{\sD} {\sD}^{c\dagger}\sD^c + \sum_{a = e, \mu, \tau} \left( M^2_{\sL a} \sL{a}^{\dagger}\sL a +  M^2_{\sE a} 
\sE{a}^{c\dag}{\sE a}^c \right)  + M^2_{H_d} H_d^{\dagger} H_d + M^2_{H_u} H_u^{\dagger} H_u\nn
&& + (M^2_{\se d} H_d^{\dag}\sLe + H.c.) -\left(B_0 H_d H_u + B_e \sLe H_u  + H.c.\right).
\end{eqnarray}
\end{widetext}

One can explicitly check that even after the addition of the above soft SUSY-breaking 
terms, the model still has the \newa symmetry.

Without going into the details of the calculations, we can make some comments based on 
symmetries. The electroweak symmetry is spontaneously broken by the vacuum expectation values (VEVs) of the two Higgs 
fields $H_u$ and $H_d$. In addition, if the sneutrino fields ${\tilde \nu}_\mu$ and ${\tilde \nu}_\tau$ acquire nonzero
VEVs then both the \newu and \newa are broken down spontaneously to nothing. Thus we have two massless Goldstone 
bosons, one of which makes the \newu gauge boson massive and the other one, the Majoron, exists in the spectrum of particles. 
This Majoron is a $CP-$odd particle and the physical
spectrum also has a very light $CP-$even scalar partner to the $CP-$odd massless Majoron\cite{Grossman:2002ry}. Hence, such a scenario is excluded as the $Z-$boson decay into the Majoron and its $CP-$even scalar partner has not been observed experimentally. We must study the scalar sector in some detail to see this explicitly.

\subsection{Scalar sector}
\label{min-scalar-sector}
As the $U(1)_{L_e}$ symmetry is explicitly broken, we cannot distinguish between 
${\hat L}_e$ and ${\hat H}_d$ superfields because all of their quantum numbers are the same. In principle the scalar components of both ${\hat L}_e$ and ${\hat H}_d$
get nonzero VEVs. We use the above freedom of 
indistinguishability to choose a basis where only one of them gets a nonzero 
VEV. 
In our subsequent discussion we shall work in a basis where the VEV of the electron sneutrino ${\tilde \nu}_e$ is rotated away. 

The total scalar potential is given by
\begin{eqnarray}
 V_{\rm scalar} =  V_F + V_D + V_{\rm soft}.
\end{eqnarray}
where $V_F$ is calculated from Eq. \ref{e:mini:superpot} using
\begin{equation}
V_F= \sum_i \left|\frac{\partial W}{\partial \Phi_i}\right|^2
\end{equation}
and
\begin{equation}
V_D= \frac{1}{2}D_a D_a + \frac{1}{2}D^2_Y + \frac{1}{2} D^2_X
\end{equation}
where $D_a= \surd 2 g_a \phi^{\ast} T^a \phi$ and $V_{soft}$ is the scalar part of Eq. \ref{e:mini:soft}.
In the supersymmetric gauged \newu model the gauge kinetic term mixing affects the gauge fields, the gauginos,
and the auxiliary fields $D_Y$ and $D_X$, where $X = L_\mu - L_\tau$. The auxiliary fields can be written, using their
equations of motion, as\cite{Suematsu:1998wm} 
\bea
&&D_Y = - \sum_i g^\prime \frac{Y^i}{2} |\phi_i|^2, \nn 
&&D_X = - \sum_i \left(g_m \frac{Y^i}{2} + g_X \frac{Q_X^i}{2}\right) |\phi_i|^2,\label{eqn:dydx}
\eea
where $Y^i$ and $Q^i_X$ are the charges of the scalar fields $\phi_i$ corresponding to 
$U(1)_Y$ and $U(1)_{L_\mu -L_\tau}$ gauge symmetry, respectively. The
gauge coupling associated with \newu is $g_X$ , while $g_m$ is the coupling
generated via kinetic mixing.

The contributions of the neutral scalar fields to the scalar potential is as 
follows:
\begin{widetext}
\begin{eqnarray}
V_{\rm neut}&=& \left(m^2_{H_u} + |\mu|^2 + |\mu_e|^2\right)|h_u^0|^2 + \left(m^2_{H_d} + |\mu|^2 
\right)|h_d^0|^2 + \sum_{a = e,\mu,\tau} \left(M^2_{\tilde L_a } + |\mu_e|^2 \delta_{e,a}
\right)|\tilde \nu_a|^2  \nn
&&+ \bigg( \mu^\ast \mu_e h_d^{0\ast}\sneue - B 
h_u^0 h_d^0  - B_e\sneue h_u^0  + M^2_{\se d} h_d^{0\ast}\sneue + 
H.c.\bigg)\nn
&&+ \frac{1}{8}\left(g_1^2+g_2^2\right)\left(|h_u^0|^2-|h_d^0|^2 - \sum_{a = e,\mu,\tau} 
|\tilde \nu_a|^2\right)^2\nn 
&&+ \frac{1}{8}\left(g_m \left(|h_u^0|^2-|h_d^0|^2- \sum_{a = e,\mu,\tau} 
|{\tilde \nu}_a|^2 \right) - g_x\left(|\sneutau |^2-|\sneumu |^2 
\right)\right)^2.
\end{eqnarray}
\end{widetext}

We assume that only the neutral scalar fields $H_u$, $H_d$, ${\tilde \nu}_\mu$ and ${\tilde \nu}_\tau$ acquire nonzero VEVs while minimizing 
the scalar potential and the VEVs are defined as $\langle h_u^0 \rangle \equiv v_u/\sqrt{2}$, $\langle h_d^0 \rangle \equiv v_d/\sqrt{2}$, 
$\langle {\tilde \nu}_\mu \rangle \equiv v_\mu/\sqrt{2}$ and $\langle {\tilde \nu}_\tau \rangle \equiv v_\tau/\sqrt{2}$.  The minimization equations are
\begin{widetext}
\begin{eqnarray}
&&B v_d + \frac{1}{8}g_x g_m (|v_\tau|^2 - |v_\mu|^2)v^\ast_u - \frac{1}{8}(g^2 + {g^\prime}^2 + g_m^2)(|v_u|^2 - |v_0|^2 )v^\ast_u
 - (M^2_{H_u} + |\mu_0|^2) v^\ast_u = 0 \\
&&B v_u - \frac{1}{8}g_x g_m (|v_\tau|^2 - |v_\mu|^2)v^\ast_d + \frac{1}{8}(g^2 + {g^\prime}^2 + g_m^2)(|v_u|^2 - |v_0|^2 )v^\ast_d
- (M^2_{H_d} + |\mu|^2) v^\ast_d = 0 \\
&&v^\ast_\mu \bigg[M^2_{{\tilde L}_\mu} - \frac{g_x^2}{8}(|v_\tau|^2 - |v_\mu|^2) + \frac{1}{8}g_x g_m (|v_u|^2 - |v_d|^2 - 2 |v_\mu|^2) - \frac{1}{8}(g^2 + {g^\prime}^2 + g_m^2)(|v_u|^2 - |v_0|^2 ) \bigg] = 0\label{min-eq-min-mod-vmu}\\
&&v^\ast_\tau \bigg[M^2_{{\tilde L}_\tau} + \frac{g_x^2}{8}(|v_\tau|^2 - |v_\mu|^2) - \frac{1}{8}g_x g_m (|v_u|^2 - |v_d|^2 - 2 |v_\tau|^2)- \frac{1}{8}(g^2 + {g^\prime}^2 + g_m^2)(|v_u|^2 - |v_0|^2 ) \bigg] = 0\label{min-eq-min-mod-vtau}\\
&&\bigg[ M^2_{\se d} + \mu^\ast \mu_e \bigg]v_d^\ast = B_e v_u
\end{eqnarray}
\end{widetext}
where $|v_0|^2 \equiv |v_d|^2 + |v_\mu|^2 + |v_\tau|^2$ and $|\mu_0|^2 \equiv |\mu|^2 + |\mu_e|^2.$

The vacuum expectation values are such that 
\bea
v \equiv (|v_u|^2 + |v_0|^2)^{1/2} = \frac{2 m_W}{g}
\eea

\subsection{Case of both $v_\mu$ and $v_\tau$ $\neq 0$}
If we demand that both $v_\mu$ and $v_\tau$  are nonzero, 
then we have two corresponding massless Goldstone bosons in the spectrum. There is 
always a Goldstone boson arising because of nonzero VEVs of the Higgs fields $H_u$ and $H_d$. Two of these 
three Goldstone bosons can be eaten up by the neutral gauge bosons $Z$ and $Z^\prime$. The remaining 
massless $CP-$odd Majoron is a physical particle and hence experimentally ruled out from the nonobservation of 
such particles in the decay of the $Z$ boson. We can understand this even better if we calculate the $CP-$even and $CP-$odd neutral scalar mass-squared matrices for these scenarios.


\subsection{Scalar mass matrices}
\label{min-scalar-mass-matrices}

We can calculate the $CP-$even and $CP-$odd neutral scalar mass-squared matrices from the $CP-$even and $CP-$odd neutral scalar potential, using
\begin{eqnarray}
M^2_{ij}= \frac{\partial^2 V}{\partial\phi_i \partial\phi_j}\bigg|_{min} \label{eqn:massformula}.
\end{eqnarray}
The $CP-$even scalar mass matrix in the basis $(h_u,h_d,\sneue,\sneumu,\sneutau)$, is given by
\begin{widetext}
\begin{eqnarray}
&&M^2_{\rm even} = \nn
&&\left(
\begin{array}{ccccc}
B v_d/v_u + \frac{1}{4}{\tilde g}^2 v^2_u & -B - \frac{1}{4}{\tilde g}^2 v_u v_d & -B_e & -\frac{1}{4} ({\tilde g}^2  - g_m g_x) v_u v_\mu 
& -\frac{1}{4} ({\tilde g}^2 + g_m g_x) v_u v_\tau \\
-B - \frac{1}{4}{\tilde g}^2 v_u v_d & {\tilde m}^2_{h_d^0 h_d^{0 \ast}} + \frac{1}{4}{\tilde g}^2 v_d^2 & {\tilde m}^2_{de} + \mu \mu_e 
& \frac{1}{4} ({\tilde g}^2 - g_m g_x) v_d v_\mu & \frac{1}{4} ({\tilde g}^2 + g_m g_x) v_d v_\tau \\
-B_e & {\tilde m}^2_{de} + \mu \mu_e & {\tilde m}^2_{{\tilde \nu}_e {\tilde \nu}^\ast_e} & 0 & 0 \\
-\frac{1}{4} ({\tilde g}^2  - g_m g_x) v_u v_\mu & \frac{1}{4} ({\tilde g}^2 - g_m g_x) v_d v_\mu & 0 & {\tilde m}^2_{{\tilde \nu}_\mu {\tilde \nu}^\ast_\mu} 
+ \frac{1}{4} {g^{\prime \prime}_-}^2 v^2_\mu & \frac{1}{4}({\tilde g}^2 - g^2_x) v_\mu v_\tau \\
-\frac{1}{4} ({\tilde g}^2 + g_m g_x) v_u v_\tau & \frac{1}{4} ({\tilde g}^2 + g_m g_x) v_d v_\tau & 0
& \frac{1}{4}({\tilde g}^2 - g^2_x) v_\mu v_\tau & {\tilde m}^2_{{\tilde \nu}_\tau {\tilde \nu}^\ast_\tau} + \frac{1}{4} {g^{\prime \prime}_+}^2 v^2_\tau 
\end{array}
\right) \nn
\label{CP-even-min-model}
\end{eqnarray}
\end{widetext}
Here ${\tilde g}^2 = (g^2 + {g^\prime}^2 + g^2_m)$ and ${g^{\prime \prime}_\mp}^2 = {\tilde g}^2 \mp 2 g_m g_x + g^2_x$,
\begin{eqnarray}
{\tilde m}^2_{h_d^0 h_d^{0 \ast}} &=& m^2_{H_d} + \mu^2 - \frac{1}{8} {\tilde g}^2 
(v^2_u - v^2_d - v^2_\mu - v^2_\tau)\nn
&&+ \frac{1}{8} g_m g_x (v^2_\tau - v^2_\mu), 
\\
{\tilde m}^2_{{\tilde \nu}_e {\tilde \nu}^\ast_e} &=& M^2_{{\tilde L}_e} + \mu_e^2 - \frac{1}{8} {\tilde g}^2 
(v^2_u - v^2_d - v^2_\mu - v^2_\tau)  \nn
&&+ \frac{1}{8} g_m g_X (v^2_\tau - v^2_\mu), 
\\
{\tilde m}^2_{{\tilde \nu}_\mu {\tilde \nu}^\ast_\mu} &=& M^2_{{\tilde L}_\mu}  - \frac{1}{8} {\tilde g}^2 
(v^2_u - v^2_d - v^2_\mu - v^2_\tau)\nn  
&&+ \frac{1}{8} g_m g_x (v^2_u - v^2_d - 2 v^2_\mu)\nn
&& - \frac{1}{8} g^2_x (v^2_\tau - v^2_\mu), 
\\
{\tilde m}^2_{{\tilde \nu}_\tau {\tilde \nu}^\ast_\tau} &=& M^2_{{\tilde L}_\tau} - \frac{1}{8} {\tilde g}^2 
(v^2_u - v^2_d - v^2_\mu - v^2_\tau) \nn 
&&- \frac{1}{8} g_m g_x (v^2_u - v^2_d - 2 v^2_\tau)\nn
&& + \frac{1}{8} g^2_x (v^2_\tau - v^2_\mu). 
\end{eqnarray}

The $CP-$odd scalar mass matrix in the basis $(h_u,h_d,\sneue,\sneumu,\sneutau)$ is given by
\begin{eqnarray}
&&M^2_{\rm odd} = \nn
&&\left(
\begin{array}{ccccc}
B v_d/v_u & B & B_e & 0& 0 \\
B & {\tilde m}^2_{h_d^0 h_d^{0 \ast}} & {\tilde m}^2_{de} + \mu \mu_e & 0 & 0 \\
B_e & {\tilde m}^2_{de} + \mu \mu_e & {\tilde m}^2_{{\tilde \nu}_e {\tilde \nu}^\ast_e} & 0 & 0 \\
0 & 0 & 0 & {\tilde m}^2_{{\tilde \nu}_\mu {\tilde \nu}^\ast_\mu} & 0 \\
0 & 0 & 0 & 0 & {\tilde m}^2_{{\tilde \nu}_\tau {\tilde \nu}^\ast_\tau} 
\end{array}
\right) \nn
\label{CP-odd-min-model}
\end{eqnarray}

When both $v_\mu$ and $v_\tau$ are nonzero, Eqs.(\ref{min-eq-min-mod-vmu}) and (\ref{min-eq-min-mod-vtau})
give us 
\bea
{\tilde m}^2_{{\tilde \nu}_\mu {\tilde \nu}^\ast_\mu} = 0 = {\tilde m}^2_{{\tilde \nu}_\tau {\tilde \nu}^\ast_\tau}
\eea
This gives two massless Goldstone bosons from the $CP-$odd mass matrix as discussed earlier. In addition, the diagonalization of 
the upper $3 \times 3$ block gives another massless Goldstone boson which is absorbed by the $Z-$boson. 

Let us now consider the $CP-$even scalar squared masses by calculating the eigenvalues of the matrix in Eq.(\ref{CP-even-min-model}).
It is straightforward to check that the eigenvector 
\bea
&&\rho = 
\frac{1}{K} \left(
\begin{array}{c}
2 v_u v_\mu v_\tau \\
2 v_d v_\mu v_\tau \\
0 \\
v_\tau(v^2_u - v^2_d) \\
v_\mu(v^2_u - v^2_d)
\end{array}
\right) ,\nn &&K = \sqrt{v^2_\mu (v^2_u -v^2_d)^2 + v^2_\tau (v^2_u -v^2_d)^2 + 4 v^2_\mu v^2_\tau (v^2_u + v^2_d)}\nn
\eea
corresponds to a zero eigenvalue of $M^2_{even}$.
This means that at the tree level there exists a massless $CP-$even scalar, $\rho$. 
However, $\rho$ gains a small mass ${\cal {O}}(\sqrt{v^2_\mu + v^2_\tau})$ when radiative corrections are incorporated since it is not a Goldstone boson. The nonobservation
of the $Z-$boson decay $Z \rightarrow ~{\rm Majoron}~ + \rho$ in experiments rules out the minimal model described above. 


\subsection{Case of either $v_\mu \neq 0$ or $v_\tau \neq 0$}

On the other hand, the problem related to the massless Majoron discussed above can be ameliorated if only one of the two sneutrinos (namely, ${\tilde \nu}_\mu$ and ${\tilde \nu}_\tau$) acquires a VEV. In this case we have two possibilities:
\begin{eqnarray}
{\rm Type A:} && v_\mu \ne 0, v_\tau =0; \nn
&&U(1)_{L_\mu-L_\tau} \times U(1)_{L_\mu+L_\tau} \to U(1)_{L_\tau}  
\label{type-A}\nn
\eea
\bea
{\rm Type B:} && v_\mu = 0, v_\tau \ne 0; \nn
&&U(1)_{L_\mu-L_\tau} \times U(1)_{L_\mu+L_\tau} \to U(1)_{L_\mu} 
\label{type-B}\nn
\end{eqnarray}
In both of these cases there is no massless Majoron in the physical spectrum and either of these two scenarios are equally viable. 

For $\tilde \nu_\kappa$ ($\kappa = \mu ~{\rm or}~ \tau$) the minimization equation (assuming all 
parameters are real) is,
\bea
&&v_\kappa \bigg[M^2_{{\tilde L}_\kappa} + \frac{g_x^2}{8}v_\kappa^2 + \frac{1}{8} Q_\kappa g_x g_m (v_u^2 - v_d^2 - 2 v_\kappa^2)\nn
&&- \frac{1}{8}(g^2 + {g^\prime}^2 + g_m^2)(v_u^2 - v_0^2 ) \bigg] = 0\nn
\label{min-eq-min-mod-vkappa}
\eea
where $Q_\kappa$ are the \newu charges corresponding to ${\tilde \nu}_\mu$ and ${\tilde \nu}_\tau$. 

From Eq. \ref{min-eq-min-mod-vkappa} we get (for $v_\kappa \neq 0$)
\begin{eqnarray}
{\tilde m}^2_{{\tilde \nu}_\kappa {\tilde \nu}^\ast_\kappa} &=&M^2_{\tilde L_\kappa} + \frac18 \left( g_1^2 +g_2^2 + g_m^2\right) \left(v_\kappa^2- v_u^2 + v_d^2 \right)\nn
&&+ \frac14 g_x^2 v_\kappa^2 + 
\frac14 Q_\kappa g_m g_x \left(v_u^2 -v_d^2 - 2 v_\kappa^2 \right)\nn
&=&0. 
\label{conditions-min-model}
\end{eqnarray}
In  the pseudoscalar mass matrix, all of the off-diagonal entries of the column and 
row corresponding to the field $\tilde \nu_\kappa$ are zero and the diagonal 
entry is nothing but ${\tilde m}^2_{{\tilde \nu}_\kappa {\tilde \nu}^\ast_\kappa}$ [see, Eq.\ref{CP-odd-min-model}]. 
Thus if we demand $v_\kappa \ne 0$, which in turn implies that the condition \ref{conditions-min-model} 
must be true, then there exists a corresponding massless pseudoscalar state as discussed in Sec. \ref{min-scalar-mass-matrices}. 
This massless pseudoscalar is eaten up by the neutral gauge field corresponding to \newu gauge symmetry. In addition, there is a 
Goldstone boson that gives mass to the $Z-$boson.  Thus, there is no massless Majoron present in the physical spectrum of this model. 

\subsection{Failure of the minimal model}\label{ss:mini.mod}
We have seen in the previous section that the models of Type A [Eq.(\ref{type-A})] and Type B [Eq.(\ref{type-B})] have residual global symmetries 
$U(1)_{L_\mu}$ and $U(1)_{L_\tau}$, respectively. Because of the presence of such global symmetries in
each type of model after the electroweak symmetry breaking, textures of the Majorana neutrino mass matrix 
[in the basis ($\nu_e$, $\nu_\mu$, $\nu_\tau$)] and the charged lepton mass matrix [in the basis ($e$, $\mu$, $\tau$)] 
should have, in general, the following forms:
\begin{eqnarray}
 {\rm Type A:} & & m_\nu = \left(\begin{array}{ccc} \checkmark & \checkmark & 0
\\ \checkmark & \checkmark & 0 \\ 0& 0& 0
\end{array}   \right), ~m_\ell = \left(\begin{array}{ccc}
\checkmark & \checkmark & 0 \\ \checkmark & \checkmark & 0 \\ 0& 0& \checkmark
\end{array}   \right), \nn
\end{eqnarray}
\begin{eqnarray}
{\rm Type B:} & & m_\nu = \left(\begin{array}{ccc} \checkmark & 0 & \checkmark
\\ 0& 0& 0
\\ \checkmark & 0  & \checkmark
\end{array}   \right), ~m_\ell = \left(\begin{array}{ccc}
\checkmark & 0  & \checkmark  \\ 0& \checkmark & 0 \\ \checkmark & 0 &
\checkmark
\end{array}   \right), \nn
\end{eqnarray}
where $\checkmark$ means nonzero entries. Note that neutrino mass matrix has one less 
nonzero entry compared to the charged lepton mass matrix because of the Majorana nature 
of the neutrinos. With the above textures of these mass matrices, the resulting Pontecorvo-Maki-Nakagawa-Sakata (PMNS) matrix will not be able 
to reproduce the correct pattern of neutrino mixing as observed in different neutrino experiments. Thus, these two models with minimal field content are ruled out in the light of neutrino experimental data.

\section{The nonminimal model}\label{sec:nonminimal}

We have seen that the minimal model is not phenomenologically attractive. The source of this problem was essentially the fact that there is either an accidental U(1)$_{L_{\mu}+L_{\tau}}$ which is broken along with \newu, or that there is a residual U(1)$_{L_{\mu}}$/U(1)$_{L_{\tau}}$ that spoils the neutrino mass matrix texture. 
The solution is to have extra fields, $\eta$ and $\etabar$, that are charged only under
\newu and couple to $\mu/\tau$, to make sure that U(1)$_{L_{\mu}+L_{\tau}}$, U(1)$_{L_{\mu}}$ and U(1)$_{L_{\tau}}$ are not symmetries of the theory.
An additional benefit is the fact that now we have fields that are singlet under all SM gauged symmetries that can acquire vacuum expectation values to spontaneously break \newu, i.e. $\langle\eta\rangle = v_{\eta}/\sqrt{2}$ and $\langle\etabar\rangle = v_{\etabar}/\sqrt{2}$. While there is no problem even if the sneutrinos do acquire VEV, we consider the situation where they do not, that is to say $\langle\sneumu\rangle =\langle\sneutau\rangle = 0$. This has more to do with simplifying the calculation than with any technical glitches, although one could argue that this minimizes tree-level $Z/Z^{\prime}$ mixing and dissociates \newu breaking from electroweak symmetry breaking. To this end we also take $g_m =0$ in subsequent calculations. This ensures that there is no mixing between $Z$ and $Z^{\prime}$ at tree level and the mass of the new gauge boson is given simply by
\begin{equation}
M^2_{Z^{\prime}} = \frac{g_X^2}{4}(v_{\eta}^2 + v_{\etabar}^2).
\end{equation}

The field content and \newu charges of the nonminimal model are shown in Table \ref{table-non-min-model-charges}.
\begin{table}[htb]
\centering
\begin{tabular}{|c|ccccccccccc|cc|}
\hline\hline
Superfields & ${\hat Q}_i$ & ${\hat U}^c_i$ & ${\hat D}^c_i$ & ${\hat L}_e$ & ${\hat E}^c_e$ & ${\hat L}_\mu$ & ${\hat E}^c_\mu$ & ${\hat L}_\tau$ & ${\hat E}^c_\tau$ &
${\hat H}_u$ & ${\hat H}_d$ & $\hat \eta$ & $\hat {\bar \eta}$ \\
\hline
$U(1)_{L_\mu - L_\tau}$ & 0 & 0 & 0 & 0 & 0 & 1 & -1 & -1 & 1 & 0 & 0 & -1 & 1 \\
\hline
\end{tabular}
\caption{$U(1)_{L_\mu - L_\tau}$ charge assignments to the chiral superfields of the nonminimal model}
\label{table-non-min-model-charges}
\end{table}

The superpotential for the above choice of charges is
\begin{eqnarray}\label{e:non-min:superpot}
W &=& W^{min}+\epsilon_{ij}\bigg[- y_\eta {\hat L}_\mu^i {\hat H}_u^j {\hat \eta} - y_{\bar \eta} {\hat L}_\tau^i {\hat H}_u^j {\hat {\bar \eta}} \bigg] + \mu_\eta {\hat \eta} {\hat {\bar \eta}}.\nn
\end{eqnarray}

Here too we have considered baryon number parity as in the minimal model. The bilinear $R$-parity-violating parameter for the first generation (i.e., $\mu_1$) in the nonminimal model, is the same as the parameter $\mu_e$ in the minimal model.
\subsection{Free from gauge anomalies}
Let us now discuss the conditions of anomaly cancellation\cite{Bilal:2008qx} in this model.
\begin{enumerate}
\item It is not required to examine the anomaly condition involving all 
possible 
combinations of $SU(3), SU(2)$ and $U(1)_{Y}$ because the MSSM is anomaly free.
\item 
The anomaly cancellation condition for $\{SU(3),SU(3)\} \,$ \newu
is satisfied as none of the colored particles are charged under \newu.
\item
The $SU(2)$ fields which are charged under \newu  are $L_{\mu}$ and 
$L_\tau$. As they have opposite charges the $\{SU(2),SU(2)\} \,$ \newu anomaly cancellation condition is also satisfied.
\item
Similarly the $\{U(1)_{Y}, U(1)_{Y} \} \,$ \newu anomaly cancellation 
condition is also satisfied because the \newu charges of $L_\mu$ and 
$E^c_\mu$ are opposite to $L_\tau$ and $E^c_\tau$ respectively.
\item
One can check that the $\{$ \newu , \newu$\} \, U(1)_{Y} $ condition is 
also satisfied:
\bea
&& \underbrace{2\times 1^2\times (-\frac12)}_{L_\mu} +  \underbrace{2\times 
(-1)^2\times (-\frac12)}_{L_\tau} \nn 
&+& \underbrace{(-1)^2\times 1}_{E^c_\mu} + 
\underbrace{1^2\times 1}_{E^c_\tau} =0.
\eea
\item
The cubic anomaly for \newu is satisfied:
\begin{equation}
\underbrace{2\times 1^3}_{L_\mu} +  \underbrace{2\times 
(-1)^3}_{L_\tau} + \underbrace{(-1)^3}_{E^c_\mu} + 
\underbrace{1^3}_{E^c_\tau} =0.
\end{equation}
\item
And finally, the mixed anomaly with gravity is also satisfied, as the trace of the 
charges of fields for this new gauge group vanishes.
\end{enumerate}
Thus, all the gauge anomalies are canceled out. The gauge anomalies pertaining to the two extra superfields $\eta$ and $\etabar$ cancel out among themselves as they are singlets under all SM gauge symmetries and oppositely charged under the \newu symmetry.

\subsection{Vacua and scalar masses}
We must consider the entire scalar potential of the model and minimize it to obtain the vacuum expectation values of the various fields. Just as in the case of the minimal model, the total scalar potential is 
\begin{equation}
V= V_F+V_D+V_{soft}\label{eqn:vtotal},
\end{equation}
where $V_F$ is calculated from Eq. \ref{e:non-min:superpot}
and $V_{soft}$ comes from the soft SUSY-breaking terms in the Lagrangian, 
\begin{eqnarray}
-\mathcal{L}_{soft}&=&-\mathcal{L}_{soft}^{min}-\bigg(A_{\eta} \eta\sLm H_u 
+ A_{\bar{\eta}} \bar{\eta}\sLt H_u + h.c\bigg)\nn
&&+ M^2_{\eta} \eta^{\dag}\eta + M^2_{\bar{\eta}}\bar{\eta}^{\dag}\bar{\eta} +\left(B_{\eta}\eta\bar{\eta} + H.c.\right).\label{eqn:lsoft}
\end{eqnarray}
\vspace{1em}

$V_D$ is calculated in exactly the same way as for the minimal model [see Eq. \ref{eqn:dydx}], including contributions from two new scalar fields $\eta$ and $\etabar$.

The neutral scalar potential,

\begin{widetext}
\begin{eqnarray}
V_{neut}&=& \left(m^2_{H_u} + |\mu|^2 + |\mu_1|^2\right)|h_u^0|^2 + \left(m^2_{H_d} + |\mu|^2 
\right)|h_d^0|^2 + \sum_{a = e,\mu,\tau} \left(M^2_{\tilde L_a } + |\mu_1|^2 \delta_{e,a}
\right)|\tilde \nu_a|^2  \nn
&&+\left(M_{\eta}^2 + \mu_{\eta}^2\right)|\eta|^2 + \left(M_{\etabar}^2 + \mu_{\eta}^2\right)|\etabar|^2 +\bigg(\frac12 y_{\eta}^2 |\sneumu|^2 |h_u^0|^2 +\frac12 y_{\eta}^2 |\sneutau|^2 |h_u^0|^2 \nn
&&+\frac{1}{2}y_{\eta}^2|\eta|^2|\sneumu|^2 + \frac{1}{2}y_{\eta}^2|\eta|^2|h_u^0|^2 +\frac{1}{2} y_{\bar{\eta}}^2|\bar{\eta}|^2|\sneutau|^2 
+  \frac{1}{2}y_{\bar{\eta}}^2|\bar{\eta}|^2|h_u^0|^2 + y_{\bar{\eta}}y_{\eta} \eta^{\ast}\bar{\eta}\sneumu^{\ast}\sneutau + H.c.\bigg) \nn
&&- \bigg(-y_{\eta}\mu \eta^{\ast}\sneumu^{\ast}h_d^0-y_{\eta}\mu_1\eta^{\ast}\sneumu^{\ast}\sneue -y_{\bar{\eta}}\mu \bar{\eta}^{\ast}\sneutau^{\ast}h_d^0 
- \mu_1 y_{\bar{\eta}}\bar{\eta}^{\ast}\sneutau^{\ast}\sneue + y_{\eta}\mu_{\eta}h_u^{0\ast}\sneumu^{\ast}\bar{\eta}+ y_{\bar{\eta}}\mu_{\eta}h_u^{0\ast}\sneutau^{\ast}\eta\nn 
&&+A_{\eta}\eta\sneumu h_u^0 + A_{\bar{\eta}}\bar{\eta}\sneutau h_u^0 
- \mu\mu_1 h_d^{0\ast}\sneue + B h_u^0 h_d^0  + B_e\sneue h_u^0 - M^2_{\se d} h_d^{0\ast}\sneue - B_{\eta}\eta\bar{\eta} + H.c.\bigg)\nn 
&&+\frac{1}{8}\big(g_1^2+g_2^2\big)\big(|h_u^0|^2-|h_d^0|^2-|\sneue |^2-|\sneumu |^2-|\sneutau |^2\big)^2\nn &&+ \frac{1}{8}\bigg(g_m\big(|h_u^0|^2-|h_d^0|^2
-|\sneue |^2-|\sneumu |^2-|\sneutau |^2)-g_X(|\eta |^2-|\bar{\eta}|^2+|\sneutau |^2-|\sneumu |^2\big)\bigg)^2,\nn
\end{eqnarray}
\end{widetext}
is used to calculate the scalar and pseudoscalar mass-squared matrices. By replacing the fields by $(\phi_R+i\phi_I)/\surd 2$ to separate out the $CP-$even and -odd parts of the potential, we obtain

\begin{widetext}
\begin{eqnarray}
V_{even}&=& \frac{1}{2}(M^2_{H_u}+\mu^2 +\mu_1^2)(h_{uR}^0)^2 + \frac{1}{2}(M^2_{H_d}+\mu^2)(h_{dR}^0)^2 + \frac{1}{2}(M_{\tilde{L}_e}^2 +\mu_1^2)\sneueR^2 \nn 
&&+ \frac{1}{2} (M_{\tilde{L}_{\mu}}^2 + \frac{1}{2}y_{\eta}^2|\eta_R|^2 + \frac{1}{2}y_{\eta}^2|h_{uR}^0|^2)|\sneumuR|^2 +\frac{1}{2}(M_{\tilde{L}_{\tau}}^2 
+ \frac{1}{2}y_{\bar{\eta}}^2|\etabar_R|^2 + \frac{1}{2}y_{\bar{\eta}}^2|h_{uR}^0|^2)|\sneutauR|^2\nn
&&+\frac{1}{2} (M_{\eta}^2 + \frac{1}{2}y_{\eta}^2|h_{uR}^0|^2 + \mu_{\eta}^2)|\eta_R|^2 +\frac{1}{2}(M_{\bar{\eta}}^2 
+ \frac{1}{2}y_{\bar{\eta}}^2|h_{uR}^0|^2 + \mu_{\eta}^2)|\etabar_R|^2\nn
&&+\frac{1}{2}y_{\eta}y_{\bar{\eta}}\etabar_R\eta_R\sneumuR\sneutauR +\frac{1}{\sqrt{2}}y_{\eta}\mu_1\eta_R\sneumuR\sneueR + \frac{1}{\sqrt{2}}y_{\eta}\mu \eta_R\sneumuR h_{dR}^0 
+ \frac{1}{\sqrt{2}}\mu_1 y_{\bar{\eta}}\etabar_R\sneutauR\sneueR\nn
&& + \frac{1}{\sqrt{2}}\mu y_{\bar{\eta}}\etabar_R\sneutauR h_{dR}^0 - \frac{1}{\sqrt{2}}y_{\eta}\mu_{\eta}h_{uR}^0\sneumuR \etabar_R-y_{\bar{\eta}}\mu_{\eta}h_{uR}^0\sneutauR \eta_R -\frac{1}{\sqrt{2}}A_{\eta}\eta_R\sneumuR h_{uR}^0 \nn
&&- \frac{1}{\sqrt{2}}A_{\bar{\eta}}\etabar_R\sneutauR h_{uR}^0 +\mu\mu_1h_{dR}^0\sneueR - Bh_{uR}^0h_{dR}^0- B_e\sneueR h_{uR}^0 + B_{\eta}\eta_R\etabar_R\nn
&& +\frac{1}{32}(g_1^2+g_2^2+g_m^2)((h_{uR}^0)^2-(h_{dR}^0)^2-\sneueR^2-\sneumuR^2-\sneutauR^2)^2\nn
&&+\frac{1}{32}g_x^2(\eta_R^2+\sneutauR^2-\etabar^2_R-\sneumuR^2)^2\nn
&&-\frac{1}{16}g_mg_x((h_{uR}^0)^2-(h_{dR}^0)^2-\sneueR^2-\sneumuR^2-
\sneutauR^2)(\eta_R^2+\sneutauR^2-\etabar^2_R-\sneumuR^2),\label{eqn:veven}\\
V_{odd}&=& V_{neut}- V_{even}. \label{eqn:vodd}
\end{eqnarray}
\end{widetext}

We can calculate the $CP-$even and $CP-$odd neutral scalar mass-squared matrices from Eqs.\ref{eqn:veven} and \ref{eqn:vodd} using Eq.\ref{eqn:massformula}.

\subsection{Minimization of the potential}

At the minima of the potential, all of the first derivatives must vanish. The first derivatives thus give us a set of equations 
that we can plug in while calculating the second derivatives. The method is to first calculate the second derivatives of $V_{even}$ 
and $V_{odd}$ then replace the fields by their respective VEVs. At the same time, one must also replace the soft masses from the equations of minimization.

The minimization equations are
\begin{widetext}
\begin{eqnarray}
&&(\mu^2 + \mu_1^2 + \mu_2^2 + \mu_3^2 + M^2_{H_u})v_u +
 \frac{(g_1^2+g_2^2+g_m^2)}{8}(v_u^2-v_d^2)v_u -
 \frac{g_mg_x}{8}(v_{\eta}^2-v_{\etabar}^2)v_u -Bv_d =0\nn
&&(\mu^2 +  M^2_{H_d})v_d -
 \frac{(g_1^2+g_2^2+g_m^2)}{8}(v_u^2-v_d^2)v_d +
 \frac{g_mg_x}{8}(v_{\eta}^2-v_{\etabar}^2)v_d -Bv_u =0\nn
&&(M^2_{\tilde{L}_e d} + \mu\mu_1)v_d -B_ev_u =0\nn
&&(\mu\mu_2 -\mu_{\eta} \mu_2t_{\beta}\cot\gamma)v_d -A_{\eta}\frac{v_{\eta}}{\sqrt{2}}v_u =0\nn
&&(\mu\mu_3 -\mu_{\eta} \mu_3t_{\beta}t_{\gamma})v_d -A_{\etabar}\frac{v_{\etabar}}{\sqrt{2}}v_u =0\nn
&&(\mu_{\eta}^2 + \mu_2^2 t_{\xi}^2 + M_{\eta}^2)v_{\eta} + \frac{g_X^2}{8}(v_{\eta}^2-v_{\etabar}^2)v_{\eta} -\frac{g_mg_X}{8}(v_u^2-v_d^2)v_{\eta} +B_{\eta}v_{\etabar} =0\nn
&&(\mu_{\eta}^2 + \mu_3^2 t_{\xi}^2 + M_{\etabar}^2)v_{\etabar} - \frac{g_X^2}{8}(v_{\eta}^2-v_{\etabar}^2)v_{\etabar} +\frac{g_mg_X}{8}(v_u^2-v_d^2)v_{\etabar} +B_{\eta}v_{\eta} =0
\end{eqnarray}
\flushleft{where,}
\begin{eqnarray}
\mu_2 = \frac{y_{\eta}v_{\eta}}{\sqrt{2}}= \frac{\sqrt{2}y_{\eta}M_{Z^{\prime}}s_{\gamma}}{g_X}&,&\quad
\mu_3 = \frac{y_{\etabar}v_{\etabar}}{\sqrt{2}}= \frac{\sqrt{2}y_{\etabar}M_{Z^{\prime}}s_{\gamma}}{g_X},\nn
\tan\xi = \frac{v_u}{v_{\eta}}= \frac{g_XM_Ws_{\beta}}{g_2M_{Z^{\prime}}s_{\gamma}}&,&\quad
\tan\gamma = \frac{v_{\eta}}{v_{\etabar}}.
\end{eqnarray}
\end{widetext}

We have used the notation where $t_{\gamma}$ and $t_{\xi}$ means $\tan\gamma$ and $\tan\xi$, respectively, and $c_{\beta}$ and $s_{\beta}$ mean $\cos\beta$ and $\sin\beta$, respectively. Henceforth, this notation will be used in all expressions. Note that the parameters $\mu_2$ and $\mu_3$ are effective bilinear $R$-parity-violating parameters corresponding to the second and third generations, respectively [See Eq.\ref{e:non-min:superpot}].

The scalar mass squared matrices are discussed further in Appendix \ref{scalarmasssq}. Full analytic expressions for the nonzero eigenvalues of the scalar mass-squared matrices are too complicated to write down under any approximations. However, we have checked for a consistent parameter space where there are no tachyonic modes in the spectra. To get a consistent nontachyonic spectra, we were required to restrict both $\mu_{\eta}$ and $B_{\eta}$ in our formalism to be negative. For almost the entire parameter space, the lightest $CP-$even Higgs has a tree-level mass close to $M_Z$, and so at the one-loop level it is possible to get a 125 GeV Higgs.

\section{Neutralino and Neutrino Masses in the nonminimal Model}\label{sec:neutrino}

The neutralino mass terms in the Lagrangian arise in this model in the basis
\begin{equation}
\psi^{0} = \begin{pmatrix}
\nu_e, & \nu_{\mu}, & \nu_{\tau}, & i\bpino, & i\bino, & i\wino, & \hd0ino, & \hu0ino, & \etaino, & \etabarino
\end{pmatrix}
\end{equation} 
as,
\begin{widetext}
\begin{equation}
\mathcal{L}=-\frac{1}{2}\psi^{0T}M_{N}\psi^0+H.c.
\end{equation}
\flushleft{where} 
\begin{eqnarray}
&M_{N}=
\left(
\begin{array}{cc}
0&M_D\\
M_D^T&M_R
\end{array}
\right)
\end{eqnarray}
\flushleft{and,}
\begin{eqnarray}
M_D=
\left(
\begin{array}{ccccccc}
0 & 0 & 0 & 0 &-\mu_1 & 0 & 0\\
0 & 0 & 0 & 0 &-\mu_2 & -\mu_2 t_{\xi} & 0\\
0& 0 & 0 & 0 & -\mu_3 & 0 & -\mu_3 t_{\xi} t_{\gamma}
\end{array}
\right),
\end{eqnarray}
\begin{eqnarray}
M_R=
\left(
\begin{array}{ccccccc}
M_0 & 0 & 0 & 0 & 0 & -M_{Z^{\prime}}s_{\gamma} & M_{Z^{\prime}}c_{\gamma}\\
0 & M_1 & 0 & -\frac{g_1}{g_2}M_W c_{\beta} & \frac{g_1}{g_2}M_W s_{\beta} & 0 & 0\\
0 & 0 & M_2 & M_W c_{\beta} & -M_W s_{\beta} & 0 & 0\\
0 & -\frac{g_1}{g_2}M_W c_{\beta} & M_W c_{\beta} & 0 & -\mu & 0 & 0\\
0 & \frac{g_1}{g_2}M_W s_{\beta} & -M_W s_{\beta} & -\mu & 0 & 0 & 0\\
-M_{Z^{\prime}}s_{\gamma} & 0 & 0 & 0 & 0 & 0 & \mu_{\eta}\\
M_{Z^{\prime}}c_{\gamma} & 0 & 0 & 0 & 0 & \mu_{\eta} & 0
\end{array}
\right).
\end{eqnarray}
\end{widetext}
From this we can calculate the effective neutrino mass matrix \cite{Mohapatra:1979ia,Schechter:1981cv,Grimus:2000vj,Altarelli:2004za},
\begin{equation}
m_{\nu}^{eff}=-M_D M_R^{-1} M_D^T.
\end{equation}
Note that in this analysis we have taken both $g_m$ (the gauge coupling arising from kinetic mixing) and $M_{10}$ (the term corresponding to the $\bino\bpino$ term in $\mathcal{L}_{soft}$) to be zero. Although the nonminimal model does not necessarily require these to be vanishing, under this approximation not only is the neutrino mass matrix much more manageable, but there is also no $Z-Z^{\prime}$ mixing at the tree level.

Now we can write the effective neutrino mass matrix,
\begin{widetext}
\begin{eqnarray}
m_{\nu}^{eff}=\frac{1}{\Lambda}
\left(
\begin{array}{ccc}
\mu_1^2 & \mu_1\mu_2 & \mu_1\mu_3 \\ & &\\
\mu_1\mu_2 & \mu_2^2\left(1-t^2_{\xi}\frac{M_{Z^{\prime}}^2c^2_{\gamma}}{M_W^2c^2_{\beta}}\frac{g_2^2d_1}{M_gd_2}\right) & \mu_2\mu_3\left(1-t^2_{\xi}\frac{M_{Z^{\prime}}^2c_{\gamma}s_{\gamma}+M_0\mu_{\eta}}{M_W^2c^2_{\beta}}\frac{g_2^2d_1}{M_gd_2}\right)\\ & & \\
\mu_1\mu_3 & \mu_2\mu_3\left(1-t^2_{\xi}\frac{M_{Z^{\prime}}^2c_{\gamma}s_{\gamma}+M_0\mu_{\eta}}{M_W^2c^2_{\beta}}\frac{g_2^2d_1}{M_gd_2}\right) & \mu_3^2 \left(1-t^2_{\xi}t^2_{\gamma}\frac{M_{Z^{\prime}}^2s^2_{\gamma}}{M_W^2c^2_{\beta}}\frac{g_2^2d_1}{M_gd_2}\right) \label{eqn:Mneueff}
\end{array}
\right)\nn
\end{eqnarray}
\flushleft{where,}
\begin{eqnarray}
d_1 = 2\mu (M_g v_u v_d -2\mu M_1 M_2)&,&\quad
d_2 = 2\mu_{\eta}(g_X^2v_{\eta}v_{\etabar} + 2 M_0 \mu_{\eta}),\nn
M_g = g_1^2 M_2 + g_2^2 M_1&,&\quad
\Lambda = \frac{g_2^2d_1}{4M_gM_W^2c^2_{\beta}}.
\end{eqnarray}
\end{widetext}
This matrix would resemble that obtained from bilinear $R$-parity violation if the second terms inside the brackets of the lower $(2\times 2)$ block were not there. That is, it would be a rank-one matrix predicting two zero eigenvalues. This would mean that we would be unable to explain neutrino masses at the tree level.

\begin{figure}[b]
\centering
\includegraphics[scale=0.65]{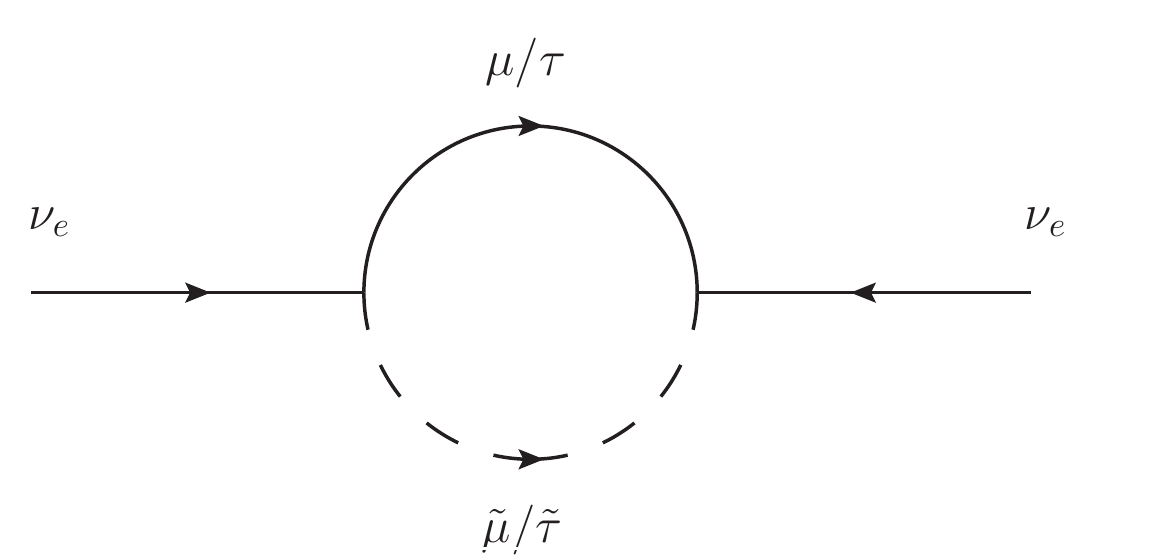}
\caption{Charged lepton-slepton loop that contributes to neutrino mass at one loop}\label{fig:numassloop}
\end{figure}

In addition to this effective light Majorana neutrino mass matrix that is generated by the seesaw effect, we have contributions to neutrino mass at the one-loop level arising from the $R$-parity-violating couplings through the diagram in Fig.\ref{fig:numassloop}.

The contribution of this diagram is given by\cite{Grossman:1998py}
\begin{eqnarray}
\left(m_{\nu}^{(1)}\right)_{11} = \sum_{p=2}^3 \frac{1}{32\pi^2} \lambda_{1pp}\lambda_{1pp} m_p \sin 2\phi_p\times \nn
 \left[-\frac{M_{p_1}^2}{m_p^2-M_{p_1}^2}\log\frac{m_p^2}{M_{p_1}^2} +
 \frac{M_{p_2}^2}{m_p^2-M_{p_2}^2}\log\frac{m_p^2}{M_{p_2}^2}\right]\label{eqn:oneloopneumass}
\end{eqnarray}
where we assume a left-right slepton mixing matrix of the form
\begin{equation}
V=\begin{pmatrix}
\cos\phi_p & \sin\phi_p \\
-\sin\phi_p & \cos\phi_p
\end{pmatrix}
\end{equation}
$M_{p_i}^2$ are the slepton mass eigenvalues, and $m_p$ are the lepton mass eigenvalues. The index $p$ denotes $\mu$ flavor when it takes the value 2 and $\tau$ when it is 3 for both sleptons and leptons. 
\begin{figure*}
\centering
\subfigure[\label{fig:cbmtest23y}]{\includegraphics[width=7.4cm]{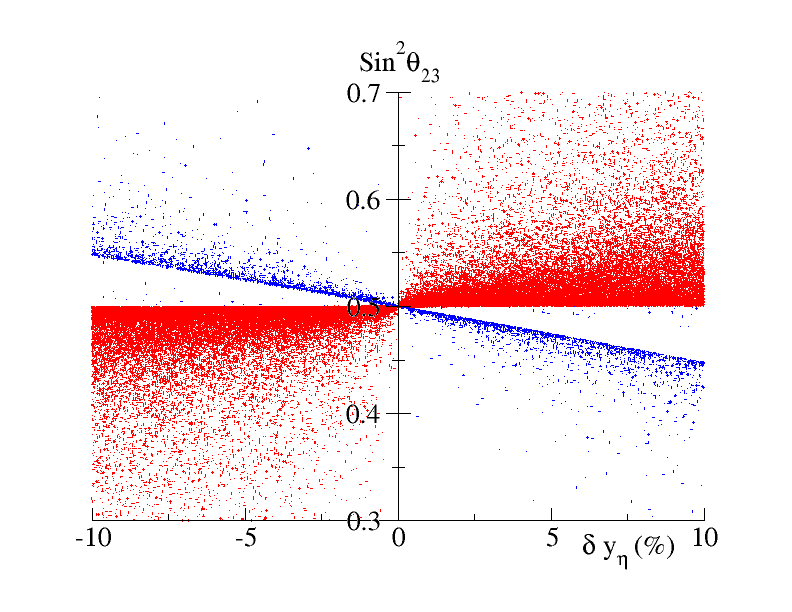}}
\subfigure[\label{fig:cbmtest13y}]{\includegraphics[width=7.4cm]{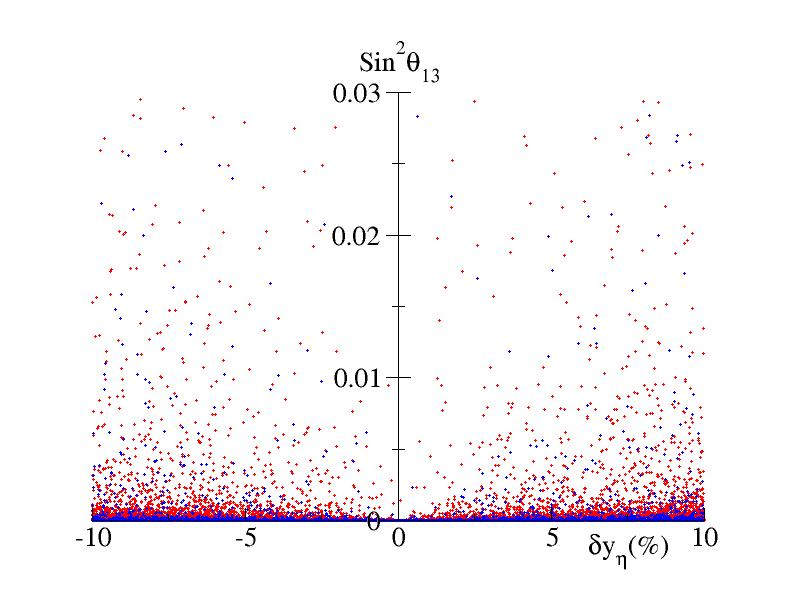}}
\subfigure[\label{fig:cbmtest23tg}]{\includegraphics[width=7.4cm]{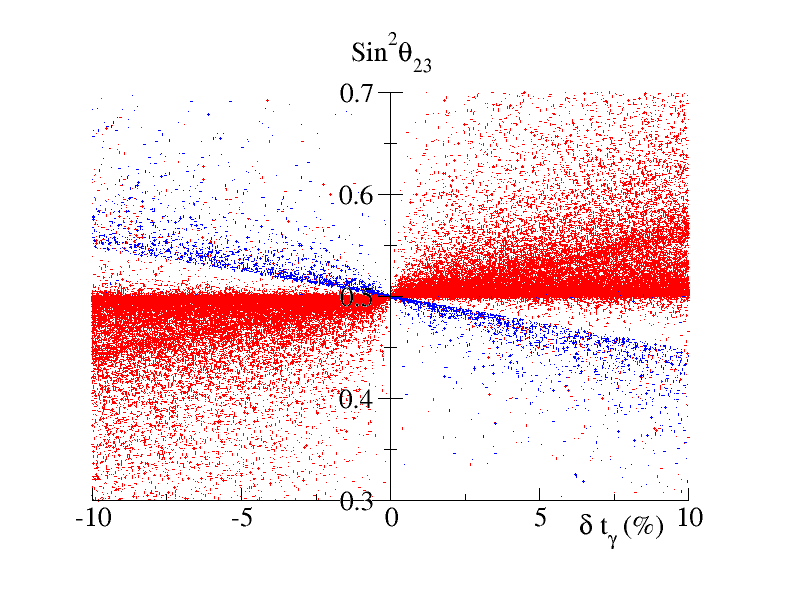}}
\subfigure[\label{fig:cbmtest13tg}]{\includegraphics[width=7.4cm]{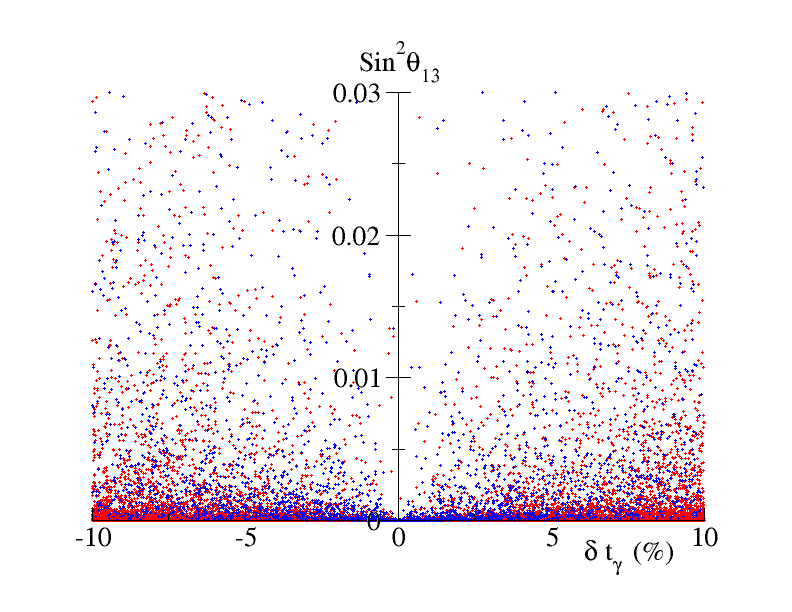}}
\caption{Variation of $\sin^2\theta_{23}$ and $\sin^2\theta_{13}$ with deviation from $y_{\eta}=y_{\etabar}$ and $\tan\gamma =1$ conditions on $m_{\nu}^{eff}$ with $\mu_{\eta}$ free. Red is for normal hierarchy, while blue is for inverted hierarchy.}\label{fig:CBtest}
\end{figure*}
A similar contribution from the quark-squark loop through the $\lambda^{\prime}$ couplings is also present in our model along with those coming from the above lepton-slepton loop. The dominant contribution in this type of diagrams come from the bottom-sbottom pair. We can ignore this contribution to the one-loop neutrino mass compared to the above contribution if we assume that the soft SUSY-breaking squark masses are higher than a few TeV. For bounds on $R$-parity-violating couplings see, for example, Ref.\cite{Dreiner:2012mx}. The one-loop corrected neutrino mass matrix is
\begin{equation}
m_{\nu}=m_{\nu}^{eff} + m_{\nu}^{(1)}.
\end{equation}

This matrix may be diagonalized by a unitary matrix $U$, such that
\begin{equation}
U m_{\nu} U^T = \text{Diag}(m_1, m_2, m_3)
\end{equation}
which is called the PMNS matrix. The most general parametrization of the PMNS matrix,
\begin{widetext}
\begin{equation}
U_{PMNS}=
\begin{pmatrix}
c_{12}c_{13} & s_{12}c_{13} & s_{13}e^{-i\delta} \\
-s_{12}c_{23}-c_{12}s_{23}s_{13}e^{i\delta} & c_{12}c_{23}-s_{12}s_{23}s_{13}e^{i\delta} & s_{23}c_{13} \\
s_{12}s_{23}-c_{12}c_{23}s_{13}e^{i\delta} & -c_{12}s_{23}-s_{12}c_{23}s_{13}e^{i\delta} & c_{23}c_{13}
\end{pmatrix}\label{eqn:PMNS}
\end{equation}
\end{widetext} 
contains three angles, $\theta_{13}$, $\theta_{12}$ and $\theta_{23}$ and the $CP$-violating phase $\delta_{CP}$. 

\subsection{Mass Models and Possible Mixing Patterns}
\begin{figure*}
\centering
\subfigure[\label{fig:neunh}]{\includegraphics[width=7cm]{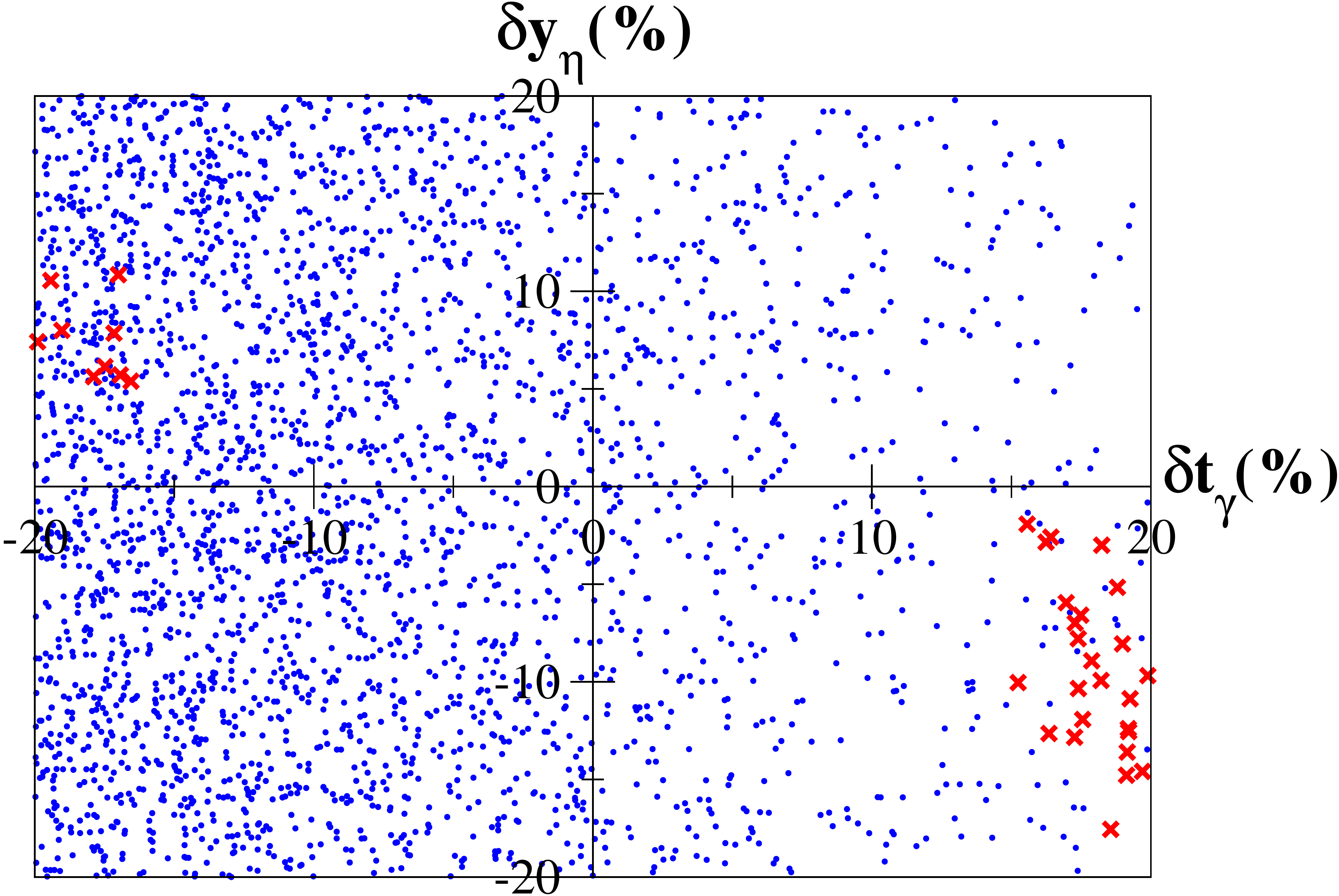}}
\subfigure[\label{fig:neuih}]{\includegraphics[width=7cm]{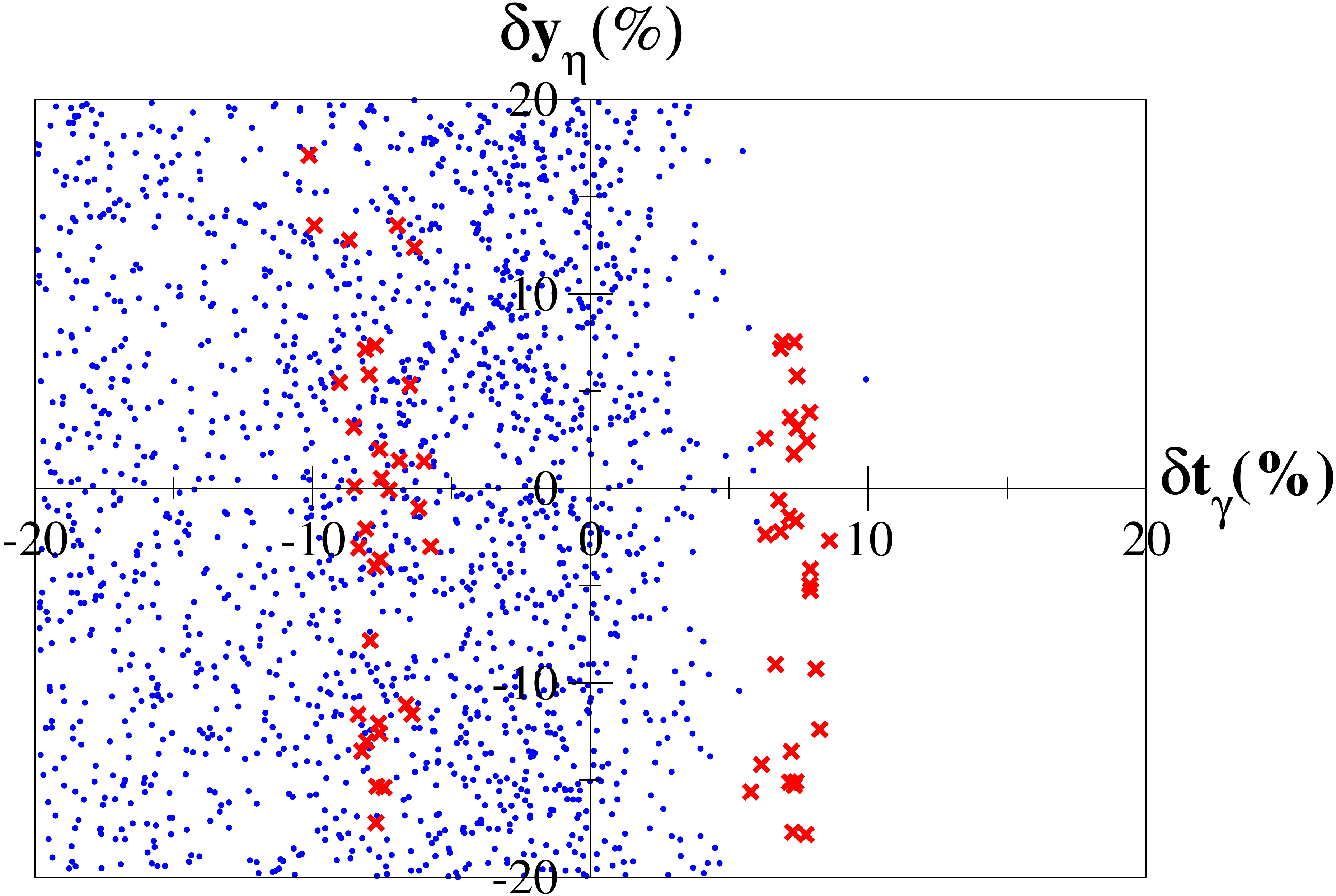}}
\caption{Points satisfying neutrino mixing angles and mass-squared differences (red) and those satisfying the muon $(g-2)$ constraint (blue) in the $\delta t_{\gamma}$ - $\delta y_{\eta}$ plane. Panel \ref{fig:neunh} is for normal hierarchy, while Panel \ref{fig:neuih} is for inverted hierarchy.}\label{fig:neuscan}
\end{figure*}
Current neutrino data favors slightly non-maximal atmospheric mixing and a nonzero $\theta_{13}$ \cite{Olive:2016xmw}. We find that, in our model, the two very simple conditions
\begin{eqnarray}
y_{\eta}&=&y_{\etabar}\nn
\tan \gamma &=& 1\label{eqn:neuconst1}
\end{eqnarray}
lead to a mass matrix of the form
\begin{equation}
M=
\left(
\begin{array}{ccccc}
a&&b&&b\\
b&&c&&d\\
b&&d&&c
\end{array}\label{eqn:mutausymmass}
\right)
\end{equation}
which is the most general $\mu-\tau$ exchange-symmetric neutrino mass matrix\cite{Grimus:2003yn}. This matrix always predicts maximal atmospheric mixing and a zero $U_{e3}$. It is by violating the conditions \ref{eqn:neuconst1} that we obtain mass matrices that satisfy neutrino oscillation data. We do not consider any $CP$-violation in our model, so $\delta_{CP}=0$ for all subsequent calculations. Our {\em modus operandi} is to compare the mixing matrices obtained with the matrix \ref{eqn:PMNS} and use
\begin{eqnarray}
\sin^2\theta_{13} &=& \lvert U_{e3} \rvert^2 \nn
\sin^2\theta_{23} &=& \frac{\lvert U_{\mu 3}\rvert^2}{1-\lvert U_{e3}\rvert^2} \nn
\sin^2\theta_{12} &=& \frac{\lvert U_{e2}\rvert^2}{1-\lvert U_{e3}\rvert^2}
\end{eqnarray}
to analyze how the mixing angles vary as we violate these conditions. We quantify the deviation from the relations (\ref{eqn:neuconst1}) by introducing two new parameters $\delta y_{\eta}$ and $\delta t_{\gamma}$,
\begin{eqnarray}
\delta t_{\gamma} &=& (1-t_{\gamma})\times 100 \nn
\delta y_{\eta} &=& \frac{y_{\eta}-y_{\etabar}}{y_{\eta}}\times 100 .
\end{eqnarray} 
These parameters are just the percentage deviation from the conditions in Eq. \ref{eqn:neuconst1}. In Fig. \ref{fig:CBtest} we plot the variation of the mixing angles with the deviation in the conditions on the Yukawa couplings, $y_{\eta}$ and $y_{\etabar}$ [see Figs. \ref{fig:cbmtest23y} and \ref{fig:cbmtest13y}] and $\tan\gamma$ [see Figs. \ref{fig:cbmtest23tg} and \ref{fig:cbmtest13tg}]. It is apparent from this figure that a variation in either of the two parameters simultaneously shifts the mixing pattern towards nonmaximal atmospheric mixing and a real, nonzero $U_{e3}$.

\subsection{Numerical analysis}

We have used MATHEMATICA 11.1 for all of our numerical analyses. For normal (NH) and inverted (IH) neutrino mass hierarchies, we found a large concentration of allowed parameter points in the regions listed in Table \ref{tab:param}.
\begin{table}
\caption{\label{tab:param}Parameter regions where large concentration of allowed points were obtained for Normal and Inverted hierarchy of neutrinos.}
\begin{ruledtabular}
\begin{tabular}{c|c|c}

&NH&IH\\
$M_{\tilde{L}}$& 0.8 to 1 TeV& 0.55 to 0.75 TeV\\

$M_1$&400 to 800 GeV&300 to 500GeV\\

$M_2$&400 to 800 GeV&1 to 1.2 TeV\\

$\mu$&200 to 300 GeV&150 to 250 GeV\\

$M_0$&50 to 100 GeV&0 to 20 GeV\\

$-\mu_{\eta}$&1 to 1.5 TeV&2 to 4 TeV\\

$M_{Z^{\prime}}$&700 to 800 GeV&1.2 to 1.5 TeV\\

$g_X$&0.4 to 0.6&0.3 to 0.4\\

$\tan\beta$&25 to 35&30 to 40\\

$y_{\eta}$,$y_{\etabar}$&$10^{-6}$ to $2\times 10^{-6}$&$3\times 10^{-6}$ to $4\times 10^{-6}$\\

$\mu_1$&$7\times 10^{-3}$ to $10^{-2}$ GeV&$2\times 10^{-3}$ to $10^{-2}$ GeV\\

$\lambda_{122}$,$\lambda_{133}$&$10^{-4}$ to $5\times 10^{-4}$&$10^{-4}$ to $2\times 10^{-4}$\\

\end{tabular}
\end{ruledtabular}
\end{table}

Here $M_{\tilde{L}}$ stands for all of the slepton soft SUSY-breaking masses. The scanned range of $M_{Z^{\prime}}$ and $g_X$ is motivated by the restrictions coming from neutrino trident production\cite{Geiregat:1990gz,Mishra:1991bv}, the LHC data from the $Z\rightarrow 4\mu$ channel\cite{Aad:2014wra,CMS:2012bw}, the observation of elastic neutrino-nucleon scattering (CE$\nu$NS) by the COHERENT Collaboration\cite{Akimov:2017ade,Shoemaker:2017lzs,Liao:2017uzy} and the observation of elastic scattering of solar neutrinos off electrons by the Borexino Collaboration \cite{Araki:2017wyg,Abdullah:2018ykz}. Apart from this the most stringent bounds on sparticle masses \cite{Aaboud:2018jiw,Patrignani:2016xqp} were also applied along with the kinematic bounds from the combined LEP data \cite{Olive:2016xmw}.

Our neutrino data consists mostly of points where the lightest neutralino is at most 6 GeV lighter than the lightest chargino and hence evades much of the constrained parameter space. 

Both of the conditions in Eq.(\ref{eqn:neuconst1}) were allowed to be violated up to 20\% and we plot the points allowed by experimental data in the $\delta t_{\gamma}$-$\delta y_{\eta}$ plane in Fig. \ref{fig:neuscan}. The points satisfying neutrino oscillation data are plotted in red while the blue background represents regions where muon $(g-2)$ is satisfied. The most stringent constraint from lepton-flavor-violating $l_j\rightarrow l_i\gamma$ processes in this model comes from $\mu\rightarrow e\gamma$ branching ratio measurements. This branching ratio never exceeds its experimental upper bound for our model in the regions where neutrino data may be satisfied\footnote{A detailed analysis of the muon anomalous magnetic moment and lepton-flavor-violating $l_j\rightarrow l_i\gamma$ processes in our model is presented later.}. 
Note that a negative deviation in $\tan\gamma$, that is, a value of $t_{\gamma}$ greater than unity, is preferred in both NH and IH from $(g-2)_{\mu}$ in these cases. 
However, this analysis is not exhaustive and there may be other regions where neutrino oscillation data may be fitted. We have only studied two interesting representative regions where we found that both neutrino and muon $(g-2)$ data are satisfied simultaneously along with all of the other aforementioned experimental bounds.

\section{Anomalous Magnetic Moment}\label{sec:madm}
The magnetic moment of the muon is one of the most accurately measured physical quantities today with the final value \cite{Olive:2016xmw}
\begin{equation}
a_{\mu}^{exp}=(116592089\pm 63)\times 10^{-11},
\end{equation}
which however does not agree with the theoretically predicted value from the Standard Model. The discrepancy,
\begin{equation}
\Delta a_{\mu}=a_{\mu}^{exp}-a_{\mu}^{SM}=(28.8\pm 8.0)\times 10^{-10},
\end{equation}
is a $\sim$3.6$\sigma $ deviation from the SM value. Given the accuracy of the $(g-2)$ measurement and the evaluation of its Standard Model prediction, it is an ideal testing 
ground for any new physics model, like SUSY. Supersymmetry, even in the MSSM has been shown to provide sizable contributions to $(g-2)$ that 
are large enough to explain its discrepancy from the SM prediction. The muon $(g-2)$ data is also ideal to constrain certain parameters of the model, such as the sign of the 
``$\mu$ term" and the mass scale of the scalar and fermionic superpartners in the case of the MSSM.

There are two main components of the MSSM contribution to the muon $(g-2)$: one is from the smuon-neutralino loop and the other is from the chargino-sneutrino loop. When the mass scales of the 
superpartners are roughly of the order of $M_{SUSY}$, this contribution is given by \cite{Chakraborty:2015bsk,Martin:2001st,Moroi:1995yh}
\begin{equation}
\Delta a_{\mu}^{MSSM}=14\text{ Sign}(\mu)\tan \beta\left(\frac{100 \text{ GeV}}{M_{SUSY}}\right)^2 10^{-10}.
\end{equation}
Our model, which has a $Z^{\prime}$ boson coupling to the muon, can complement the SUSY contribution. This allows us to have a natural solution to the hierarchy problem 
and get a stable Higgs mass, while still explaining the anomalous magnetic moment of the muon. Note that the contributions of $W$ and $Z$ bosons to the muon $(g-2)$ anomaly in our 
model are subdominant compared to the contributions mentioned above. 

\subsection{Outline of the calculation}
In our model, we have non-trivial mixing between the smuons and other charged scalars, as well as between the muons and other charged fermions. Otherwise the calculation is 
relatively straightforward and mimics that for the MSSM. Instead of the neutralino-smuon loop we consider the more general neutralino-charged scalar loops to allow for 
the mixing between smuons and other scalars. Similarly the chargino-sneutrino loop for the MSSM is expanded into a chargino-neutral scalar loop calculation. We allow the sign of the neutralino mass eigenvalues ($\epsilon_i$) and the chargino mass eigenvalues ($\eta_i$) to be either positive or negative. The diagonalizing 
matrices are suitably defined following the prescription in Appendix A of Ref.\cite{Gunion:1984yn}.

\subsubsection{Neutralino-charged scalar loop}

\begin{figure}[h]
\centering
\includegraphics[scale=0.65]{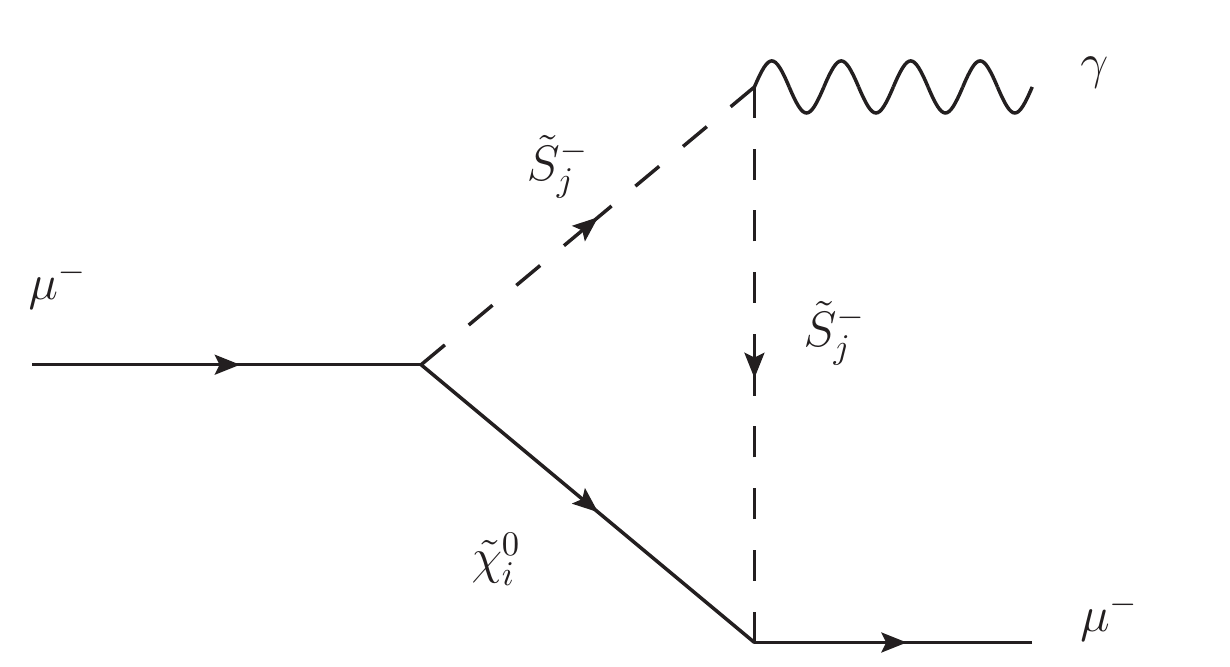}
\caption{Neutralino-charged scalar loop that contributes to muon ($g-2$)}\label{fig:ncs}
\end{figure}
For this calculation we require the neutralino mass matrix and the charged scalar mass matrix. In the basis ($\nu_e, \nu_{\mu}, \nu_{\tau}, i\bpino, i\bino, i\wino, \hd0ino, \hu0ino, \etaino, \etabarino$), we can write the neutralino mass terms as
\begin{equation}
\mathcal{L}=-\frac{1}{2}\psi^{0T}M_N\psi^0+H.c.
\end{equation}
which is diagonalized by the matrix $N$,
\begin{equation}
N^{\ast}M_NN^{\dag}=m_{\tilde{\chi}^0}.
\end{equation}
The charged scalar mass matrix ($M_{\tilde{S}^{\pm}}^2$) is written in the basis ($h_u^{+\ast}, h_d^-, \se, \smu, \stau, \seR, \smuR, \stauR$) and diagonalized so that,
\begin{equation}
U_{\tilde{S}^{\pm}}M_{\tilde{S}^{\pm}}^2U_{\tilde{S}^{\pm}}^{\dag}=m_{\tilde{S}^{\pm}}^2
\end{equation}
which includes a Goldstone mode. More about the charged scalar mass-squared matrix is discussed in Appendix \ref{csmass}.

Using these mixing matrices, the neutralino-charged scalar loop (See Fig. \ref{fig:ncs}) contribution to the muon $(g-2)$ is\cite{Chakraborty:2015bsk,Martin:2001st,Moroi:1995yh},
\begin{eqnarray}
a_{\mu}^{\chi^0}=&&-\frac{m_{\mu}}{16\pi^2}\sum_{i=1}^{10}\sum_{j=1}^{7}\bigg[\big(|n^L_{ij}|^2+|n^R_{ij}|^2\big)\frac{m_{\mu}}{12{m_{\tilde{S}^{\pm}}^2}_j}F_1^N(x_{ij})\nonumber\\
&&+\frac{{m_{\tilde{\chi}^0}}_i}{3{m_{\tilde{S}^{\pm}}^2}_j}\text{Real}\big(n^L_{ij}n^R_{ij}\big)F_2^N(x_{ij})\bigg]
\end{eqnarray}
where,
\begin{eqnarray}
F_1^N(x)&=&\frac{2(1-6x+3x^2+2x^3-6x^2\log x)}{(1-x)^4}\nonumber\\
F_2^N(x)&=&\frac{3(1-x^2+2x\log x)}{(1-x)^3}
\end{eqnarray}
with
\begin{equation*}
x_{ij}=\frac{{m_{\tilde{\chi}^0}}_i^2}{{m_{\tilde{S}^{\pm}}^2}_j}
\end{equation*}
and 
\begin{eqnarray}
n^L_{ij}=&&\left(\frac{g_X}{\surd 2}N^{\ast}_{i4}-\surd 2 g_1 N^{\ast}_{i5}-\surd 2 g_m N^{\ast}_{i4}\right){U^{\ast}_{\tilde{S}^{\pm}}}_{j7} \nn
&& -\left(y_{\mu}N^{\ast}_{i7} +\lambda_{122}N^{\ast}_{i1}\right){U^{\ast}_{\tilde{S}^{\pm}}}_{j4}\nn
&& + \lambda_{122}N^{\ast}_{i2}{U^{\ast}_{\tilde{S}^{\pm}}}_{j3}+y_{\mu}N^{\ast}_{i2}{U^{\ast}_{\tilde{S}^{\pm}}}_{j2}\\
n^R_{ij}=&&\left(\frac{g_1}{\surd 2}N^{\ast}_{i5}+\frac{g_2}{\surd 2}N^{\ast}_{i6}+\frac{g_m}{\surd 2}N^{\ast}_{i4}
-\frac{g_X}{\surd 2}N^{\ast}_{i4}\right){U^{\ast}_{\tilde{S}^{\pm}}}_{j4}\nn
&&-\left(y_{\mu}N^{\ast}_{i7}+\lambda_{122}N^{\ast}_{i1}\right){U^{\ast}_{\tilde{S}^{\pm}}}_{j7}-y_{\eta}N^{\ast}_{i9}{U^{\ast}_{\tilde{S}^{\pm}}}_{j1}.
\end{eqnarray}
In our case the external muons also mix with other charged fermions in the chargino mass matrix hence the expressions for the couplings ($n^L$ and $n^R$) will include 
appropriate elements from the chargino mixing matrices ($V_{44}$ and $U^{\ast}_{44}$ respectively). The most general formulas are given here, where $g_m$ is also nonzero. 
We take this to be zero in our numerical analysis.

\subsubsection{Chargino-neutral scalar loop}

\begin{figure}[h]
\centering
\includegraphics[scale=0.65]{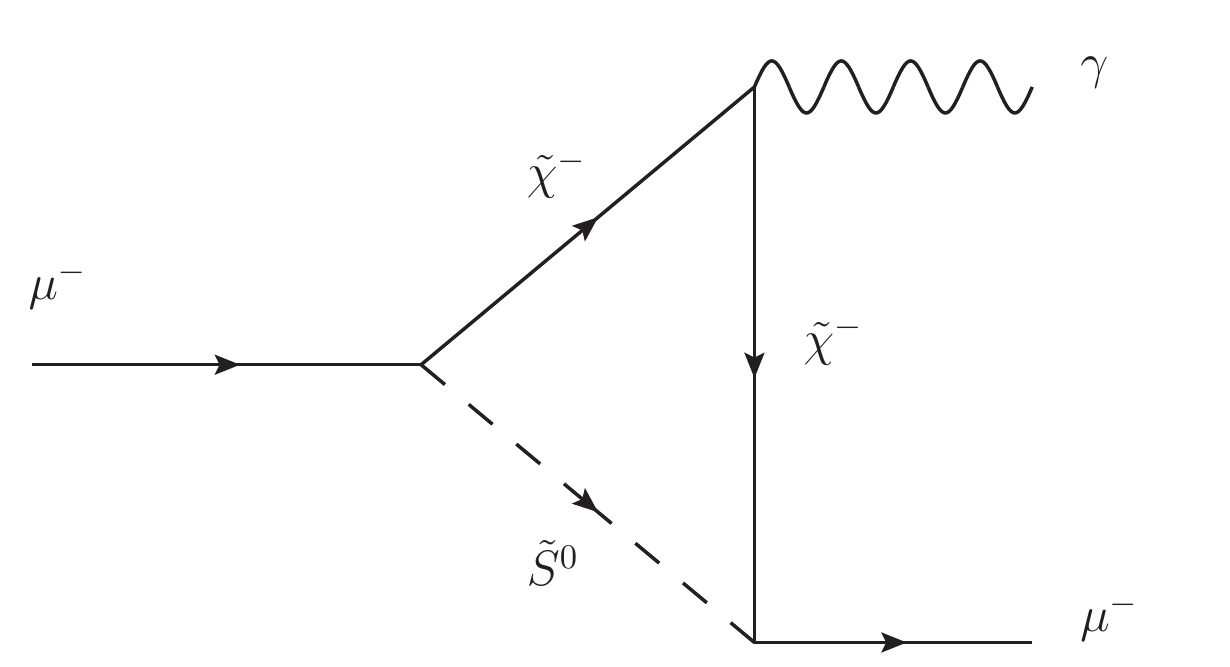}
\caption{Chargino-neutral scalar loop that contributes to muon ($g-2$)}\label{fig:cns}
\end{figure}

In this part of the calculation we require the chargino mass matrix and the neutral scalar and pseudoscalar mass-squared matrices. Defining
\begin{eqnarray}
\psi^-=(i\wmino, \hdmino, e_L^-, \mu_L^-, \tau_L^-)\nn
\psi^+=(i\wpino, \hupino, e_R^+, \mu_R^+, \tau_R^+)\nonumber
\end{eqnarray}
the chargino mass terms in the Lagrangian may be written as
\begin{equation}
\mathcal{L}\supset -\frac{1}{2}(\psi^{-T}X\psi^+ + \psi^{+T}X^T\psi^-)+h.c.
\end{equation}
where $X$ is the chargino mass matrix. It can be diagonalized by two matrices $U$ and $V$ so that,
\begin{equation}
U^{\ast}XV^{\dag}=m_{{\tilde{\chi}}^{\pm}}.
\end{equation}
The chargino mass matrix is given and discussed in Appendix \ref{chargino}.
The neutral scalar mass-squared matrix $M_{\tilde{S}^0}^2$, given in Appendix \ref{cpevenmass}, is written in the basis
\newline ($\sneueR,\sneumuR,\sneutauR,h_{dR}^0,h_{uR}^0,\eta_R,\bar{\eta}_R$) and is diagonalized so that
\begin{equation}
U_{\tilde{S}^0}^{\dag}M_{\tilde{S}^0}^2U_{\tilde{S}^0}=m_{\tilde{S}^0}^2.
\end{equation}
Similarly, the pseudoscalar mass-squared matrix $M_{\tilde{P}^0}^2$ from Appendix \ref{cpoddmass} is written in the basis 
($\sneueI,\sneumuI,\sneutauI,h_{dI}^0,h_{uI}^0,\eta_I,\bar{\eta}_I$) and is diagonalized so that
\begin{equation}
U_{\tilde{P}^0}^{\dag}M_{\tilde{P}^0}^2U_{\tilde{P}^0}=m_{\tilde{P}^0}^2.
\end{equation}
Using these mixing matrices, we calculate the contribution of the chargino-neutral scalar loop (See Fig. \ref{fig:cns}) to the muon ($g-2$)\cite{Chakraborty:2015bsk,Martin:2001st,Moroi:1995yh}
\begin{eqnarray}
a_{\mu}^{{\tilde{\chi}}^{\pm}}=&&\frac{m_{\mu}}{16\pi^2}\sum_{i=1}^5\sum_{j=1}^7\bigg[\frac{m_{\mu}}{12m_{\tilde{S}_j^0}^2}\big(|c^{Le}_{ij}|^2
+|c^{Re}_{ij}|^2\big)F_1^C(y^e_{ij}) \nn
&& +\frac{2m_{{\tilde{\chi}}_i^{\pm}}}{3m_{\tilde{S}_j^0}^2}\text{Real}\big(c^{Le}_{ij}c^{Re}_{ij}\big)F_2^C(y^e_{ij})\bigg] \nn
&&+\frac{m_{\mu}}{16\pi^2}\sum_{i=1}^5\sum_{j=1}^5\bigg[\frac{m_{\mu}}{12m_{\tilde{P}_j^0}^2}\big(|c^{Lo}_{ij}|^2+|c^{Ro}_{ij}|^2\big)F_1^C(y^o_{ij}) \nn
&&+\frac{2m_{{\tilde{\chi}}_i^{\pm}}}{3m_{\tilde{P}_j^0}^2}\text{Real}\big(c^{Lo}_{ij}c^{Ro}_{ij}\big)F_2^C(y^o_{ij})\bigg]
\end{eqnarray}
where
\begin{eqnarray}
F_1^C(x)=&&\frac{2(2+3x-6x^2+x^3+6x\log x)}{(1-x)^4}\nn
F_2^C(x)=&&-\frac{3(3-4x+x^2+2\log x)}{2(1-x)^3}
\end{eqnarray}
with
\begin{equation*}
y^e_{ij}=\frac{m_{{\tilde{\chi}}_i^{\pm}}^2}{m_{\tilde{S}_j^0}^2}, \quad y^o_{ij}=\frac{m_{{\tilde{\chi}}_i^{\pm}}^2}{m_{\tilde{P}_j^0}^2}
\end{equation*}
and
\begin{eqnarray}
c^{Le}_{ij}=&&\frac{y_{\mu}}{\sqrt{2}}U_{i2}U^{\ast}_{\tilde{S}^0_{j2}} - \frac{\lambda_{122}}{\sqrt{2}} U_{i4}U^{\ast}_{\tilde{S}^0_{j1}} + \frac{\lambda_{122}}{\sqrt{2}}U_{i3}U^{\ast}_{\tilde{S}^0_{j2}}\nn
c^{Re}_{ij}=&&-\frac{g_2}{\sqrt{2}}V_{i1}U^{\ast}_{\tilde{S}^0_{j2}} - \frac{y_{\eta}}{\sqrt{2}}V_{i2}U^{\ast}_{\tilde{S}^0_{j6}} - \frac{\lambda_{122}}{\sqrt{2}}V_{i4}U^{\ast}_{\tilde{S}^0_{j1}}\nn
c^{Lo}_{ij}=&&i\frac{y_{\mu}}{\sqrt{2}}U_{i2}U^{\ast}_{\tilde{P}^0_{j2}} - i\frac{\lambda_{122}}{\sqrt{2}} U_{i4}U^{\ast}_{\tilde{P}^0_{j1}} + i\frac{\lambda_{122}}{\sqrt{2}}U_{i3}U^{\ast}_{\tilde{P}^0_{j2}}\nn
c^{Ro}_{ij}=&&-i\frac{g_2}{\sqrt{2}}V_{i1}U^{\ast}_{\tilde{P}^0_{j2}} + i\frac{y_{\eta}}{\sqrt{2}}V_{i2}U^{\ast}_{\tilde{P}^0_{j6}} + i\frac{\lambda_{122}}{\sqrt{2}}V_{i4}U^{\ast}_{\tilde{P}^0_{j1}}. \nn
\end{eqnarray}

Just as in the case of the neutralino-charged scalar loop, here too the external muons will mix with the other charged fermions and result in factors of $V_{44}$ and 
$U_{44}^{\ast}$ in $c^L$ and $c^R$, respectively.

\subsubsection{$Z^{\prime}$ contribution}

\begin{figure}[H]
\centering
\includegraphics[scale=0.65]{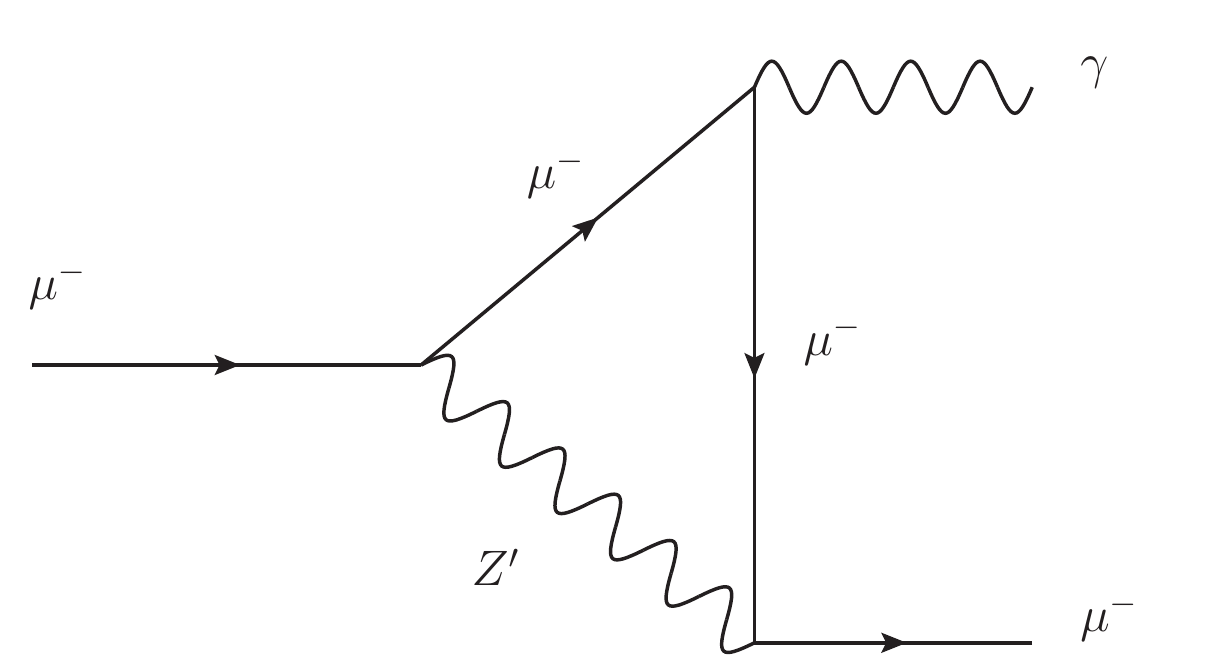}
\caption{$Z^{\prime}$ loop that contributes to muon ($g-2$)}\label{fig:zprimeloop}
\end{figure}

In addition to the purely supersymmetric contribution to $\Delta a_{\mu}$, the $Z^{\prime}$ boson also adds an important part to the total muon magnetic moment.
The contribution of \newu to the muon $(g-2)$ can be easily evaluated from the diagram in Fig. \ref{fig:zprimeloop}. It is given by\cite{Baek:2001kca,Ma:2001md,Heeck:2011wj}
\begin{equation}
\Delta a_{\mu}^{Z^{\prime}} = \frac{g_X^2m_{\mu}^2}{4\pi^2}\int_0^1 dz \frac{z^2(1-z)}{m_{\mu}^2z+M_{Z^{\prime}}^2(1-z)}.  
\end{equation}
Here too, the external muons and those inside the loop will mix with other leptons and charginos as in the previous sections. This calculation assumes no $Z-Z^{\prime}$ 
mixing at the tree level owing to the fact that $g_m$ is zero and the sneutrinos do not acquire any VEVs.

\subsection{Numerical analysis}\label{numerical_analysis}
\begin{figure*}
\centering
\subfigure[\label{fig:g-2more}]{\includegraphics[width=7cm]{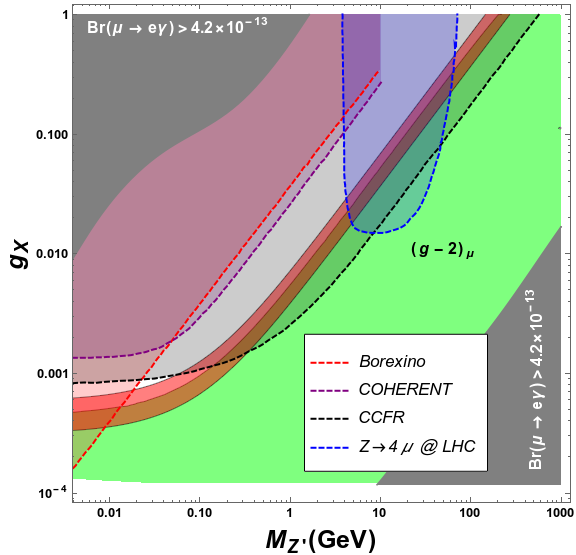}}
\subfigure[\label{fig:g-2less}]{\includegraphics[width=7cm]{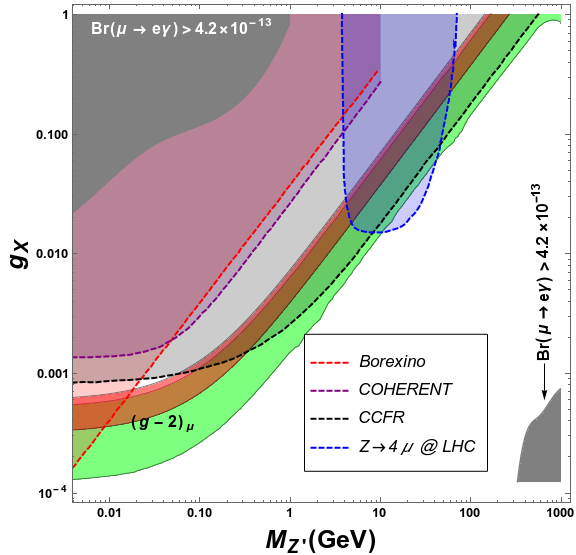}}
\caption{Parameter space for the $Z^{\prime}$ gauge boson showing the regions relevant to $(g-2)_{\mu}$. The red region is for the contribution from gauged \newu without 
considering SUSY , while the green region corresponds to our model. The brown region is the overlap. The dashed lines denote respective exclusion limits:purple for COHERENT 
neutrino elastic scattering experiment, red for the data from Borexino,  black for CCFR data for neutrino trident observation and blue for $Z\rightarrow 4\mu$ data from the
LHC. The CCFR and Z$\rightarrow$4$\mu$ exclusion regions have been taken from Ref\cite{Altmannshofer:2014pba}, while the Borexino and COHERENT exclusion regions are from 
Ref\cite{Abdullah:2018ykz}. Panel (a) represents the scenario where there is a large SUSY contribution, as opposed to Panel (b) where it 
is comparatively lower. The grey regions are ruled out from $\mu\rightarrow e\gamma$ branching ratio measurements\cite{Tanabashi:2018}.}\label{fig:MADMallowed}
\end{figure*}
Any gauged \newu model is severely constrained by neutrino trident production, that is, the production of a $\mu^+\mu^-$ pair from the scattering of a muon 
neutrino off heavy nuclei. The CHARM-II\cite{Geiregat:1990gz} and CCFR\cite{Mishra:1991bv} collaborations found reasonable agreement between the observed cross section for this 
process and its SM prediction:
\begin{eqnarray}
\frac{\sigma_{\text{CHARM-II}}}{\sigma_{\text{SM}}}=1.58\pm 0.57,\quad
\frac{\sigma_{\text{CCFR}}}{\sigma_{\text{SM}}}=0.82\pm 0.28.
\end{eqnarray} 
Thus, it severely constrains the allowed parameter space for any new neutral gauge boson. In particular, when coupled with the restrictions from the LHC data from 
the $Z\rightarrow 4\mu$ channel\cite{Aad:2014wra,CMS:2012bw}, the observation of CE$\nu$NS by the COHERENT 
Collaboration\cite{Akimov:2017ade,Shoemaker:2017lzs,Liao:2017uzy}, and the observation of elastic scattering of solar neutrinos by the 
Borexino Collaboration\cite{Araki:2017wyg,Abdullah:2018ykz}, almost the entire parameter space relevant to muon $(g-2)$ is ruled out. However, the situation for the SUSY version is not 
so bleak when it comes to resolving muon $(g-2)$ through an extra force. In our model, the total contribution to muon ($g-2$) from the two supersymmetric processes when 
added to that from the $Z^{\prime}$ loop allows for a much more liberal parameter space. 

We plot the region allowed by current $(g-2)_{\mu}$ data in the $M_{Z^{\prime}}$-$g_X$ plane for two different scenarios in fig. \ref{fig:MADMallowed}. The green
region shows the allowed parameter space in our model, while the red region shows the parameter space for a gauged \newu model where SUSY plays no part. The dashed lines show the various exclusion limits from the different experiments. The red dashed line is for the Borexino experiment (elastic scattering of solar neutrinos), while the purple 
dashed line is from the data for elastic neutrino nucleon scattering from the COHERENT Collaboration\cite{Abdullah:2018ykz}. The black dashed line shows the constraint from neutrino trident 
observations by the CCFR Collaboration\cite{Altmannshofer:2014pba}. The blue dashed line shows the exclusion limit from the LHC data of the process $Z\rightarrow4\mu$\cite{CMS:2012bw,Aad:2014wra,Altmannshofer:2014pba}. 

Figure \ref{fig:g-2more} corresponds to $M_{\smu}$=$M_{\smuR}$=500 GeV, $M_0$=70 GeV, $M_1$=400 GeV, 
$M_2$=800 GeV, $\mu$= 400 GeV and $\tan\beta$=35. Figure \ref{fig:g-2less} corresponds to $M_{\smu}$=$M_{\smuR}$=935 GeV, $M_0$=100 GeV, $M_1$=450 GeV, $M_2$=650 GeV, 
$\mu$=400 GeV and $\tan\beta$=33.5.  The rest of the SUSY parameters have been chosen judiciously for both plots: $\tan\gamma$=1.1, $\mu_1$=0.008 GeV, $\mu_{\eta}$=-3 TeV, 
$y_{\eta}$/$y_{\etabar}$=3 $\times 10^{-6}$ and the RPV $\lambda$ couplings are fixed at $10^{-4}$. The two plots were chosen to represent two different regions with differing 
magnitudes of the SUSY contribution to muon ($g-2$). Figure \ref{fig:g-2more} represents the scenario where there is a large SUSY contribution as opposed to 
Fig. \ref{fig:g-2less} where it is comparatively lower and both the SUSY and $Z^{\prime}$ contributions are by themselves insufficient to explain the anomalous magnetic 
moment of the muon. In addition, the grey regions are ruled out from measurements of $l_j\rightarrow l_i\gamma$ branching ratios. The strongest constraint comes from the 
Br($\mu\rightarrow e\gamma$) measurements while the other branching ratios are always much smaller than the current upper bounds for our choice of $y_{\eta}$/$y_{\etabar}$. The parameter space where the contribution 
from $Z^{\prime}$ dominates ($M_{Z^{\prime}}<$ 1 GeV) is already ruled out, and hence it is the SUSY contribution that we need to consider carefully. It is very clear from 
these plots that large regions of the $M_{Z^{\prime}}-g_X$ plane open up in terms of $(g-2)_{\mu}$ while the non-SUSY \newu models are already almost ruled out. 
More importantly, as we increase the SUSY contribution, the $(g-2)_{\mu}$ allowed region fills up the unconstrained parameter space in the $M_{Z^{\prime}}-g_X$ plane. 
Of course, how far we can push the SUSY contribution is limited by the Br($\mu\rightarrow e\gamma$) measurements. A larger SUSY contribution to the muon $(g-2)$ anomaly also entails 
a larger branching ratio for $l_j\rightarrow l_i\gamma$ processes. This applies constraints to a hitherto unconstrained region in the $M_{Z^{\prime}}-g_X$ plane 
($M_{Z^{\prime}}>$ 10 GeV). A detailed calculation of these branching ratios was given in Sec. \ref{sec:lfv} and Appendix \ref{lfvapp}.
\begin{figure*}
\centering
\subfigure[\label{fig:g-2neut}]{\includegraphics[width=7.0cm]{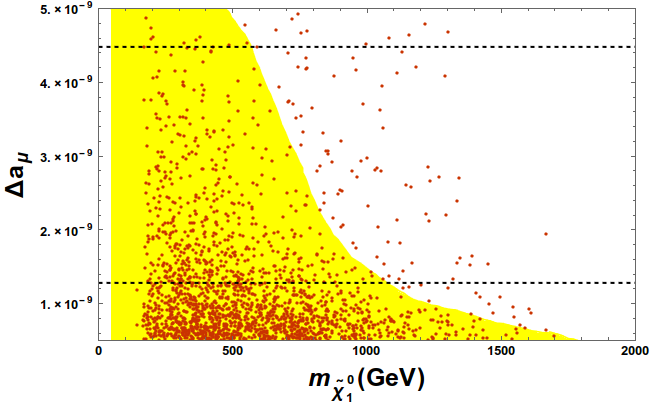}}
\subfigure[\label{fig:g-2charg}]{\includegraphics[width=7.0cm]{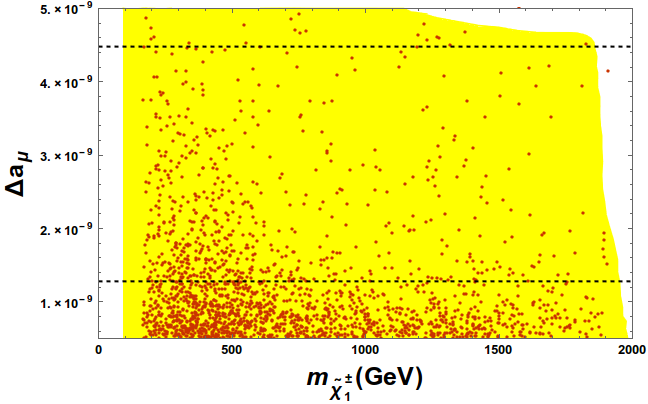}}
\subfigure[\label{fig:g-2slept}]{\includegraphics[width=7.0cm]{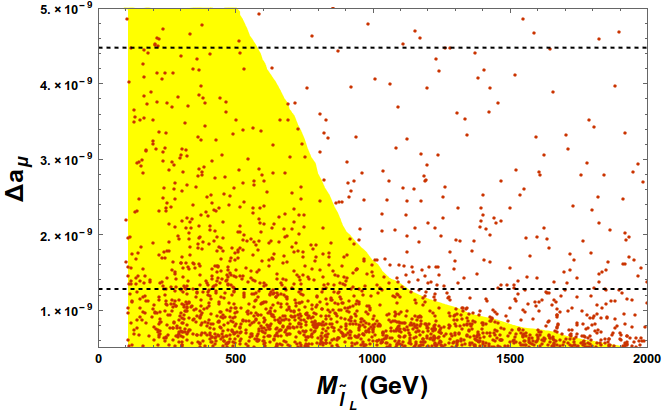}}
\caption{$\Delta a_{\mu}$ plotted against the lightest neutralino and chargino masses, along with the left-handed slepton soft mass. The 2$\sigma$ allowed region for 
$\Delta a_{\mu}$ is shown between the dashed lines. The yellow region is the allowed region from the MSSM (taken from 
Ref\cite{Kobakhidze:2016mdx})}\label{fig:MADMallowedSUSY}
\end{figure*}
In a second analysis, plot the $(g-2)_{\mu}$ against the physical masses of the lightest neutralino and chargino and the slepton soft SUSY-breaking mass in 
Fig. \ref{fig:MADMallowedSUSY}. The SUSY parameters that affect our analysis were scanned randomly in the region
\begin{eqnarray}
\text{100 GeV}< M_{\tilde{L}} <\text{2 TeV},& 
\text{100 GeV}< \mu, M_{1,2} <\text{2 TeV},\nn
\text{1 GeV}< M_{Z^{\prime}}<\text{1.5 TeV},& 
0.01<g_X<1,\nn
10^{-6}<y_{\eta}<5\times 10^{-6},& 
-\text{10 TeV}<\mu_{\eta}<-\text{1 TeV},\nn
\text{0 GeV} <M_0< \text{2 TeV},& 
\text{0 GeV}< A_l <\text{2 TeV},\nn
\text{100 GeV}< B, B_e<\text{2 TeV},&
10<\tan\beta <50.\nonumber
\end{eqnarray}

The conditions of $y_{\eta}-y_{\etabar}$ equality and $t_{\gamma}=1$ were allowed to be violated by up to 20\% and the trilinear RPV couplings were allowed to run 
from $10^{-4}$ to $5\times 10^{-4}$ while $\mu_1$ was allowed to run from $5\times 10^{-3}$ to $0.02$. The most stringent bounds on the sparticles from the latest data sets were applied along with the model-independent kinematic 
constraints on the sparticle masses from the combined result of the four LEP collaborations, just as it was done for all previous analyses.
The corresponding region allowed in the MSSM and constrained only by LEP data is shown in yellow in the same plots\cite{Kobakhidze:2016mdx}. We find that the 
$Z^{\prime}$ contribution and the SUSY contribution complement each other so that we can have heavier sparticle masses than we could in the MSSM while still explaining 
$(g-2)_{\mu}$. The approximate nondecoupling behavior that is observed is due to the extra contribution coming from the $Z^{\prime}$ loop. We have separately checked that 
the SUSY contribution alone shows the typical decoupling behavior, as expected. However, it still allows for a heavier particle spectra than can be afforded in pure MSSM. 
We have shown the data considering the most stringent sparticle limits\cite{Aaboud:2018jiw,Patrignani:2016xqp,Olive:2016xmw}. This comes from the $3l$ final-state searches 
at the LHC in chargino-neutralino pair production with slepton-mediated decays. We have also obtained similar data sets considering more relaxed bounds--the $2l$ final-state 
searches and just the LEP bounds--where we can also satisfy muon ($g-2$) for heavier sparticle masses compared to the MSSM.

\section{$l_j \rightarrow l_i \gamma$ Flavor-Violating Processes}\label{sec:lfv}

\begin{figure*}
\centering
\subfigure{\includegraphics[scale=0.65]{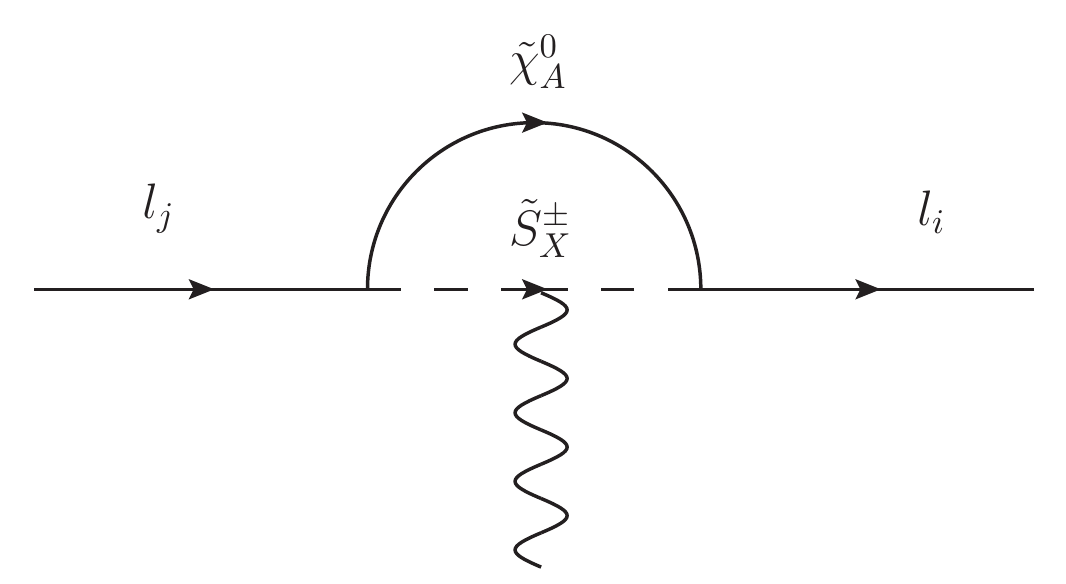}}
\subfigure{\includegraphics[scale=0.65]{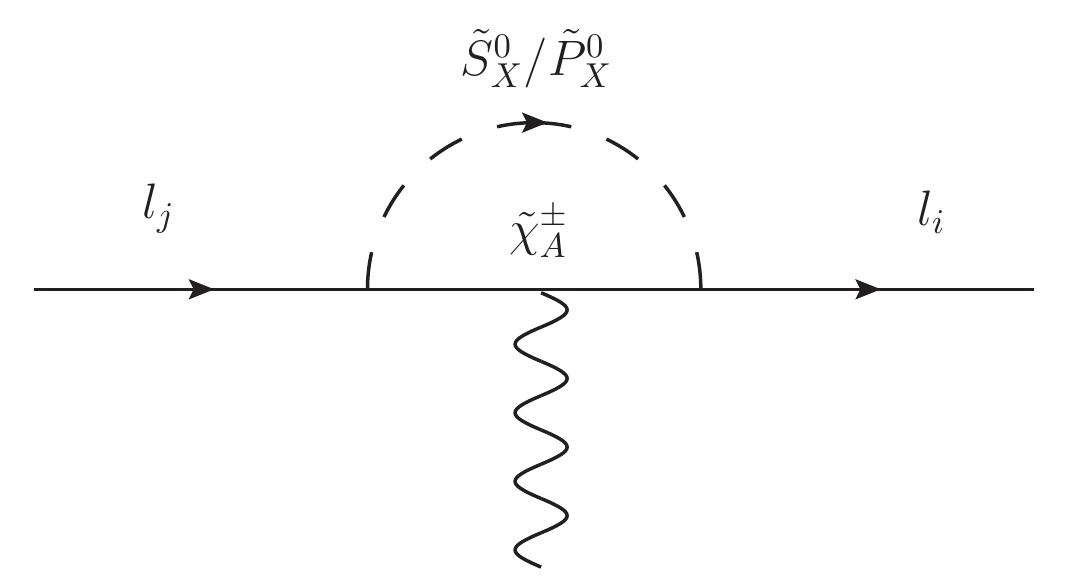}}
\subfigure{\includegraphics[scale=0.65]{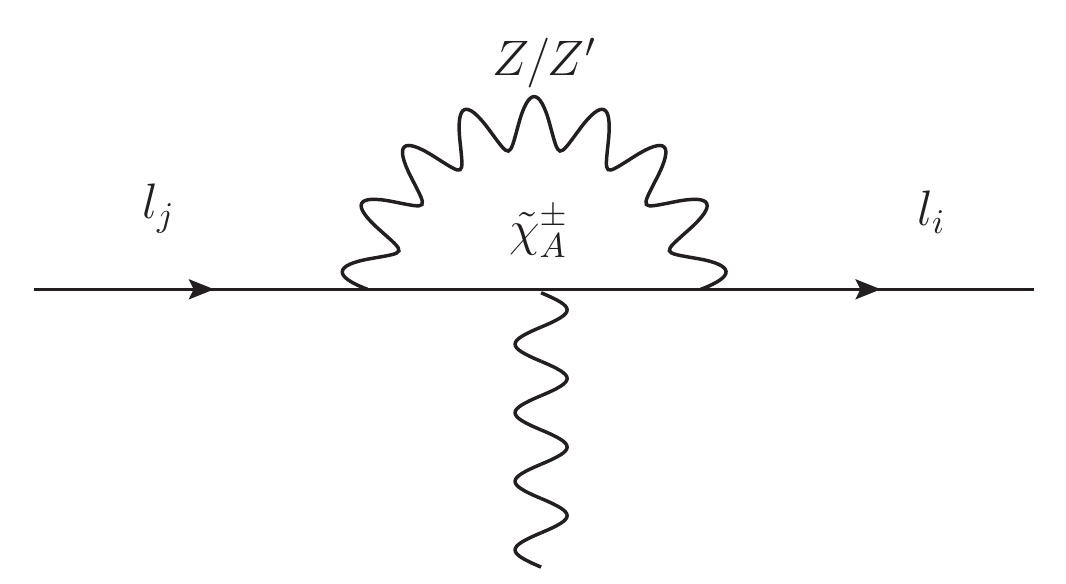}}
\subfigure{\includegraphics[scale=0.65]{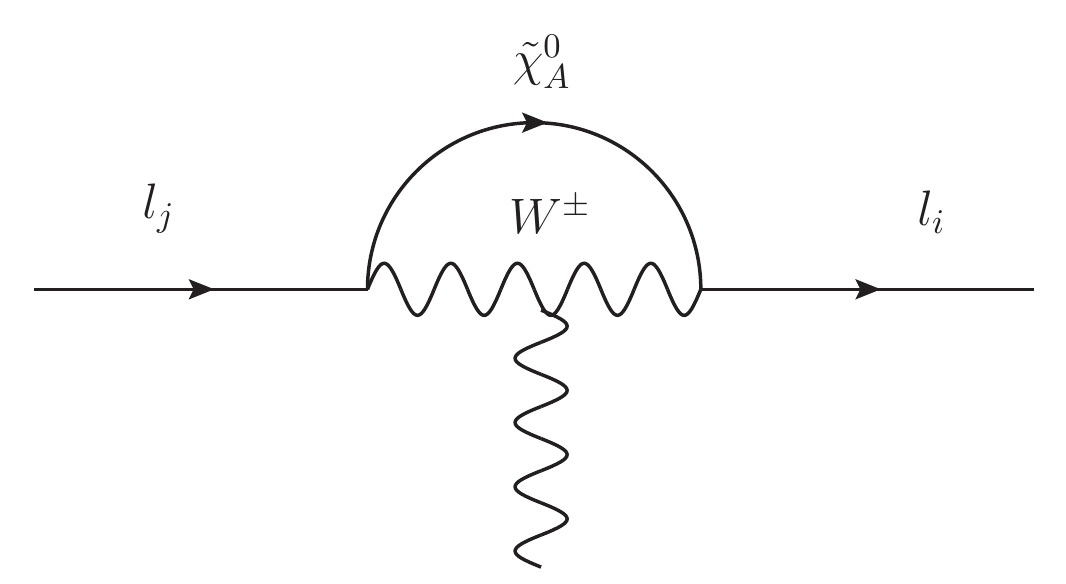}}
\caption{Diagrams contributing to $l_j \rightarrow l_i \gamma$ processes.}\label{fig:wzlfv}
\end{figure*}

Following the conventions of Sec. \ref{sec:madm} and the calculations of Ref.\cite{Carvalho:2002bq}, we can calculate the branching fractions of the lepton-flavor-violating processes $l_j \rightarrow l_i \gamma$ from the effective Lagrangian
\bea
{\cal L}_{eff} = e \frac{m_{l_j}}{2} {\bar l}_i \sigma_{\mu \nu} F^{\mu \nu}(A^L_{ij} P_L + A^R_{ij} P_R) l_j.
\eea
The Feynman diagrams contributing to these processes are given in Fig. \ref{fig:wzlfv}. The contributions involving 
squarks in the loops have been neglected as we assume their masses to be larger than a few TeV. The branching ratio for these processes is given by,
\begin{equation}
\text{BR}\left[l_j\rightarrow l_i\gamma\right]= \frac{48\pi^3\alpha}{G_F^2}\left(\left\vert A^L_{ij}\right\vert^2 + \left\vert A^R_{ij}\right\vert^2\right).
\end{equation} 
The amplitudes $A^L_{ij}$ and $A^R_{ij}$ are the left- and right-handed components respectively, of the sum of contributions from all of these diagrams, namely,  
chargino-neutral scalars, neutralino-charged scalars, $Z$-chargino, $W$-neutralino and $Z^{\prime}$-chargino fields circulating in the loop:
\begin{eqnarray}
A^L_{ij} &=& A^{L,\tilde{\chi}^0 \tilde{S}^{\pm}}_{ij} +A^{L,\tilde{\chi}^{\pm}\tilde{S}^0}_{ij} + A^{L,\tilde{\chi}^{\pm}\tilde{P}^0}_{ij} \nn
&+& A^{L,Z\tilde{\chi}^\pm}_{ij} + A^{L,Z^{\prime}\tilde{\chi}^\pm}_{ij} + A^{L,W\tilde{\chi}^0}_{ij}\nn
A^R_{ij} &=& A^L_{ij} (L/R \rightarrow R/L).
\end{eqnarray}
Detailed expressions for the amplitudes and the couplings have been relegated to Appendix \ref{lfvapp}.

The current experimental upper bounds on the branching ratios of these processes at 90\% C.L. are\cite{Tanabashi:2018},
\begin{eqnarray}
\text{BR}[\mu\rightarrow e\gamma]& < & 4.2\times 10^{-13}\nn
\text{BR}[\tau\rightarrow e\gamma]& < & 3.3\times 10^{-8}\nn
\text{BR}[\tau\rightarrow \mu\gamma]& < & 4.4\times 10^{-8}.
\end{eqnarray}

The contribution of the right-handed amplitudes $A^R$ always dominates the branching ratios over the $A^L$ for all of the processes. There are no direct couplings in our model 
that can lead to these processes, which are therefore proportional to mixing matrix elements connecting the lepton flavors. These mixings are controlled by the RPV couplings
in our model. There are two distinct scenarios where the branching ratio BR($\mu \rightarrow e \gamma$) does become large enough to have been observed by experiments.
The first case is when the neutralino-charged scalar and chargino-neutral scalar contributions dominate over the rest of the amplitudes. In this case, the contributions are 
directly proportional to and controlled by the magnitudes of the bilinear $R$-parity violating parameter $\mu_i$ appearing in our model, either explicitly or effectively when 
$\eta$/$\etabar$ acquire VEVs. There is, however, a second scenario unique to this class of models: the contribution from the $Z^{\prime}$ gauge boson. This contribution is 
controlled by the mass of the new gauge boson and the magnitude of its coupling constant g$_X$. Large contributions to BR($\mu \rightarrow e \gamma$) from these two sources, resulting in the total branching ratio exceeding $4.2\times 10^{-13}$, 
occupy two different corners of the $M_{Z^{\prime}}$-$g_X$ plane (see Fig.\ref{fig:MADMallowed} and related text in Sec. \ref{numerical_analysis}). The first process 
is dominant when the mass of the gauge boson is large and the coupling is small, i.e. when the effective parameters $\mu_2$ and $\mu_3$ are large. This is in sharp contrast to 
the second case which becomes pertinent only when the mass of the gauge boson is below 1 GeV and the branching ratio increases with increasing $g_X$. 

Still, neither of these situations really matter when it comes to a viable parameter space that can explain both neutrinos and the muon magnetic moment simultaneously. To this 
end we must note that the bilinear RPV parameters that control the branching ratios play a vital role in neutrino mass generation. Similarly, the $Z^{\prime}$ loop 
contributing to these processes also makes up for the most important source of the muon anomalous magnetic moment in our model. As a result, large contributions to 
$l_j\rightarrow l_i \gamma$ branching ratios from the fermion-scalar loops become significant only when the $\mu_i$ parameters are too large to accommodate neutrino masses 
below 0.1 eV. On the other hand, when the contribution of the additional gauge boson to BR($\mu\rightarrow e\gamma$) is larger than the current experimental upper bound, 
its contribution to the muon anomalous magnetic moment is also too large. The amplitudes pertaining to $W^\pm$ and $Z$ are always negligible 
compared to the three sources of lepton-flavor violation discussed above.

\section{Conclusion}\label{sec:conc}
We began with an attempt to explore whether gauged \newu extended SUSY could help improve the current situation when it comes to neutrino oscillation data 
and the muon anomalous magnetic moment given the current experimental bounds on SUSY itself. The minimal model with the MSSM field content required the sneutrinos 
to acquire nonzero vacuum expectation values so that the gauged \newu is spontaneously broken. This lead to two different problems. First, when both the 
sneutrinos charged under the new gauged symmetry acquired VEVs the model suffered from the Majoron problem wherein we had a massless $CP-$odd scalar and its light 
$CP-$even partner which could couple to the $Z$-Boson. Second, even when only one of the sneutrinos acquired a VEV, we found that the neutrino mass matrix had a texture 
that was impossible to fit to current oscillation data. Hence the minimal model was ruled out and the nonminimal model was adopted. 

Now, there are two extra fields that are singlets under all other gauge symmetries except \newu and acquire VEVs to spontaneously break the symmetry instead of 
the sneutrinos. This allowed us to avoid the Majoron problem altogether. We found extremely intriguing results when it came to neutrino mixing. Under two very 
simple assumptions, the model resulted in the most $\mu-\tau$ symmetric mass matrix. These conditions were that the two new fields acquire VEVs of equal magnitude 
and sign, and that the two new Yukawa couplings be equal. Of course, this mass matrix yielded maximal atmospheric mixing and a zero $\theta_{13}$ which is ruled 
out by current data. So we parametrized deviations from this exact mixing pattern with two parameters for the two conditions and this allowed us to obtain multiple 
regions where the correct neutrino oscillation could be explained. Deviations in the Yukawa coupling equality complemented that in the equality of the VEVs in the sense 
that the former alone is insufficient to fit neutrino data but is often necessary in conjunction with the latter.

In parallel, we conducted a numerical analysis of the muon anomalous magnetic moment in our model. A scan over the entire parameter space showed that we can explain 
the observed magnetic moment of the muon for much larger values of the sparticle masses compared to the MSSM. We also observed an intriguing nondecoupling behavior in 
the plots of $\Delta a_{\mu}$ vs sparticle masses owing to the presence of the $Z^{\prime}$ boson which can make up for any decrease in the SUSY contribution with heavier 
sparticle masses. As expected, decoupling was still observed if we did not consider the $Z^{\prime}$ contribution. Finally, we combined both of these analyses to show two 
representative regions of the parameter space (Fig. \ref{fig:neuscan}) where neutrino oscillation data may be reconciled with muon magnetic moment measurements. It 
is possible to fit neutrino oscillation data with both normal and inverted hierarchies of the masses. In the plot showing these regions, we superposed the regions explaining 
$(g-2)_{\mu}$, which further restricted the parameter space in both cases. Still, it is possible to obtain a parameter space where the model explains neutrino oscillation 
data along with the muon anomalous magnetic moment.

Last, we conducted a full study of the charged lepton-flavor-violating processes $l_j\rightarrow l_i\gamma$ in our model. We found that there are two possible situations 
where the branching ratio BR($\mu \rightarrow e \gamma$) may constrain our model. However, these regions are either already ruled out by other experiments and result in a 
prediction of $(g-2)_{\mu}$ that is too large, or they give rise to neutrino masses that are too large to accommodate current oscillation data. We included the contours of the 
constrained region from these processes in the plots wherever they were pertinent. We conclude that in any region of physical interest, the branching ratios are predicted 
in our model to be too small to have been detected by experiments to date.

Some interesting signatures for this model at the LHC would be the three or more leptons plus missing energy final state, involving supersymmetric particles in the 
intermediate states. 
For example, pp$\rightarrow \tilde{\mu}^+\tilde{\mu}^-Z^{\prime}$/$\tilde{\nu}_{\mu}\tilde{\nu}_{\mu}^{\ast}Z^{\prime}$/$\tilde{\nu}_{\mu}\tilde{\mu}^+Z^{\prime}$ 
processes can lead to multilepton final states along with $\mET$. In addition, contributions to multilepton final states (with or without $\mET$) 
involving SM particles and $Z^{\prime}$ can also be present. A detailed analysis with all possible final states requires a separate dedicated study, which we plan to 
undertake in a future work.

\begin{acknowledgments}
We thank  Aritra Biswas, Sabyasachi Chakraborty and Avirup Shaw for helpful discussions. S.R. acknowledges the workshop, ``Blueprints Beyond the Standard Model" organized 
at the Tata Institute of Fundamental Research, Mumbai and ``NuHorizons VII" organized at Harish-Chandra Research Institute, Allahabad for important insights.
\end{acknowledgments}

\appendix
\section{Scalar Mass-Squared Matrices}\label{scalarmasssq}
\subsection{$CP-$even mass squared matrix}\label{cpevenmass}
The $CP-$even mass-squared matrix was derived using Eq. \ref{eqn:massformula} in the basis,\newline $(\sneueR,\sneumuR,\sneutauR,h_{dR}^0,h_{uR}^0,\eta_R,\bar{\eta}_R)$. We assume that $g_m$ is zero. The mass matrix may be expressed as
\begin{eqnarray}
M_{\tilde{S}^0}^2=B_{\rm even}+\tilde{M}^2_{\rm even}\label{eqn:nseven}
\end{eqnarray}
where
\begin{widetext}
\begin{eqnarray}
B_{\rm even}=&&\nn
&&\left(
\begin{array}{ccccccc}
\mu_1^2 & \mu_1\mu_2 & \mu_1 \mu_3 & B_e t_{\beta} & -B_e & 0 & 0\\
\mu_1\mu_2 & \mu_2^2 (1+t_{\xi}^2) & \mu_2\mu_3 & \mu \mu_2 & -\mu\mu_2 \cot\beta & \mu_{\eta}\mu_2t_{\xi}\cot\gamma & -\mu_{\eta}\mu_2t_{\xi}\\
\mu_1\mu_3 & \mu_2\mu_3 & \mu_3^2(1+t^2_{\xi}t^2_{\gamma}) & \mu\mu_3 & -\mu\mu_3\cot\beta & -\mu_{\eta}\mu_3t_{\xi}t_{\gamma} & \mu_{\eta}\mu_3t_{\xi}t^2_{\gamma}\\
B_et_{\beta} & \mu\mu_2 & \mu\mu_3 & Bt_{\beta} & -B & 0 & 0\\
-B_e & -\mu\mu_2\cot\beta & \mu\mu_3\cot\beta & -B & B\cot\beta & 2\mu_2^2t_{\xi} & 2\mu_3^2 t_{\xi}t_{\gamma}\\
0 & \mu_{\eta}\mu_2t_{\xi}\cot\gamma & -\mu_{\eta}\mu_3t_{\xi}t_{\gamma} & 0 & 2\mu_2^2t_{\xi} & -B_{\eta}\cot\gamma & B_{\eta}\\
0 & -\mu_{\eta}\mu_2t_{\xi} & \mu_{\eta}\mu_3t_{\xi}t^2_{\gamma} & 0 & 2\mu_3^2t_{\xi}t_{\gamma} & B_{\eta} & -B_{\eta}t_{\gamma}\\
\end{array}
\right) \nn
\nn
\nn
%
\tilde{M}^2_{\rm even}=&&\diag\big(M_{\tilde{L}_e}^2 + \tilde{m}^2_1,M_{\tilde{L}_{\mu}}^2 + \tilde{m}^2_2,M_{\tilde{L}_{\tau}}^2 + \tilde{m}^2_3,(M_Z)_{2\times 2},(M_{Z^{\prime}})_{2\times 2}\big)
\end{eqnarray}
\end{widetext}
and 
\begin{eqnarray}
(M_Z)_{2\times 2} &=& \begin{pmatrix}
M_Z^2c^2_{\beta} & -M_Z^2c_{\beta}s_{\beta}\\
-M_Z^2c_{\beta}s_{\beta} & M_Z^2s^2_{\beta}
\end{pmatrix}\nn
(M_{Z^{\prime}})_{2\times 2} &=& \begin{pmatrix}
M_{Z^{\prime}}^2s^2_{\gamma} & -M_{Z^{\prime}}^2s_{\gamma}c_{\gamma}\\
-M_{Z^{\prime}}^2s_{\gamma}c_{\gamma} & M_{Z^{\prime}}^2c^2_{\gamma}
\end{pmatrix}
\end{eqnarray}
where all the parameters are as defined for the minimization equations and $\tilde{m}^2_i= \frac{M_Z^2}{2}c_{2\beta}+ Q_X^i M_{Z^{\prime}}^2c_{2\gamma}$.
\subsection{$CP$-Odd mass-squared matrix}\label{cpoddmass}

The $CP-$odd mass-squared matrix is constructed by taking second derivatives according to Eq.\ref{eqn:massformula} in the basis 
$(\sneueI,\sneumuI,\sneutauI,h_{dI}^0,h_{uI}^0,\eta_I,\bar{\eta}_I)$. The symmetric matrix is given by
\begin{widetext}
\begin{equation}
M_{\tilde{P}^0}^2=B_{\rm odd}+\tilde{M}^2_{\rm odd}\label{eqn:nsodd}
\end{equation}
\flushleft{where} 
\begin{eqnarray}
B_{\rm odd}=&&\nn
&&\left(
\begin{array}{ccccccc}
\mu_1^2 & \mu_1\mu_2 & \mu_1 \mu_3 & B_e t_{\beta} & B_e & 0 & 0\\
\mu_1\mu_2 & \mu_2^2 (1+t_{\xi}^2) & \mu_2\mu_3 & \mu \mu_2 & -\mu\mu_2 \cot\beta & -\mu_{\eta}\mu_2t_{\xi}\cot\gamma & -\mu_{\eta}\mu_2t_{\xi}\\
\mu_1\mu_3 & \mu_2\mu_3 & \mu_3^2(1+t^2_{\xi}t^2_{\gamma}) & \mu\mu_3 & \mu\mu_3\cot\beta & -\mu_{\eta}\mu_3t_{\xi}t_{\gamma} & -\mu_{\eta}\mu_3t_{\xi}t^2_{\gamma}\\
B_et_{\beta} & \mu\mu_2 & \mu\mu_3 & Bt_{\beta} & B & 0 & 0\\
B_e & -\mu\mu_2\cot\beta & \mu\mu_3\cot\beta & B & B\cot\beta & 2\mu_2^2t_{\xi} & 2\mu_3^2 t_{\xi}t_{\gamma}\\
0 & -\mu_{\eta}\mu_2t_{\xi}\cot\gamma & -\mu_{\eta}\mu_3t_{\xi}t_{\gamma} & 0 & 2\mu_2^2t_{\xi} & -B_{\eta}\cot\gamma & -B_{\eta}\\
0 & -\mu_{\eta}\mu_2t_{\xi} & -\mu_{\eta}\mu_3t_{\xi}t^2_{\gamma} & 0 & 2\mu_3^2t_{\xi}t_{\gamma} & -B_{\eta} & -B_{\eta}t_{\gamma}\\
\end{array}
\right) \nn
\nn
\nn
\tilde{M}^2_{\rm odd}=&&\diag\left(M_{\tilde{L}_e}^2 +\tilde{m}^2_1, M_{\tilde{L}_{\mu}}^2 +\tilde{m}^2_2, M_{\tilde{L}_{\tau}}^2 +\tilde{m}^2_3,0,0,0,0\right).
\end{eqnarray}
\end{widetext}
Again, all of the parameters are as defined for the minimization equations. There are two exactly zero eigenvalues of this mass matrix that correspond to the two Goldstone modes arising from the spontaneous breaking of the gauge symmetries to U(1)$_{em}$. The corresponding matrix in the MSSM has just one zero eigenvalue. The extra Goldstone mode corresponds to the breaking of \newu and gives mass to the $Z^{\prime}$ boson.

\subsection{Charged scalar mass-squared matrix}\label{csmass}
The charged scalar mass-squared matrix is calculated from the total potential \ref{eqn:vtotal} using Eq. \ref{eqn:massformula} in the basis 
$u^+=(h_u^+,h_d^{-\dag},\se^{\dag},\smu^{\dag},\stau^{\dag},\sEe,\sEm,\sEt)^T$ and $u^-=(h_u^{+\dag},h_d^{-},\se,\smu,\stau,\sEe^{\dag},\sEm^{\dag},\sEt^{\dag})^T$. 
The respective terms may be written down as
\begin{widetext}
\begin{equation}
u^{+T}M_{\tilde{S}^{\pm}}^2u^{-}
\end{equation}
\flushleft{where,}
\begin{equation}
M_{\tilde{S}^{\pm}}^2=B_{\pm}+\tilde{M}^2_{\pm}\label{eqn:csm}
\end{equation}
\flushleft{and,}
\begin{eqnarray}
B_{\pm}=&&\nn &&\left(
\begin{array}{cccccccc}
B \cot\beta + & B + & B_e & \mu\mu_2\cot\beta & \mu\mu_3\cot\beta & m_e\mu_1 & m_{\mu}\mu_2 & m_{\tau}\mu_3\\
M_W^2c^2_{\beta}& M_W^2c_{\beta}s_{\beta} & & & & & &\\
B + & B t_{\beta} + & B_et_{\beta} & \mu\mu_2 & \mu\mu_3 & m_e\mu_1t_{\beta} & m_{\mu}\mu_2 t_{\beta} & m_{\tau}\mu_3t_{\beta}\\
M_W^2c_{\beta}s_{\beta}& M_W^2s^2_{\beta} & & & & & &\\
B_e & B_e t_{\beta} & \mu_1^2 + m_e^2 & \mu_1\mu_2 & \mu_1\mu_3 & m_eX_t & \mu_2\mu_{\lambda_2}t_{\beta} & \mu_3\mu_{\lambda_3}t_{\beta}\\
& & & & & & &\\
\mu\mu_2\cot\beta & \mu\mu_2 & \mu_1\mu_2 &  \mu_2^2 + m_{\mu}^2 & \mu_2\mu_3 & 0 & m_{\mu}X_t & 0 \\
& & & & & &-\mu_1\mu_{\lambda_2}t_{\beta} &\\
\mu\mu_3\cot\beta & \mu\mu_3 & \mu_1\mu_2 & \mu_2\mu_3 & \mu_3^2 + m_{\tau}^2 & 0 & 0 & m_{\tau}X_t\\
& & & & & & &-\mu_1\mu_{\lambda_3}t_{\beta}\\
m_e\mu_1 & m_e\mu_1 t_{\beta} & m_e X_t & 0 & 0 & m_e^2 & 0 & 0 \\
& & & & & & &\\
m_{\mu}\mu_2 & m_{\mu}\mu_2 t_{\beta} & \mu_2\mu_{\lambda_2}t_{\beta} & m_{\mu}X_t& 0 & 0 & m_{\mu}^2 & 0 \\
& & & -\mu_1\mu_{\lambda_2}t_{\beta} & & & &\\
m_{\tau}\mu_3 & m_{\tau}\mu_3 t_{\beta} & \mu_3\mu_{\lambda_3}t_{\beta}& 0 &  m_{\tau}X_t& 0 & 0 & m_{\tau}^2\\
& & & & -\mu_1\mu_{\lambda_3}t_{\beta} & & &\\
\end{array}
\right) \nn 
 \nn
  \nn
\tilde{M}^2_{\pm}=&&\diag\big(0,0,\quad M_{\tilde{L}_e}^2 + \tilde{l}_{L1},\quad M_{\tilde{L}_{\mu}}^2  + \tilde{l}_{L2},\quad M_{\tilde{L}_{\tau}}^2  +\tilde{l}_{L3},
 M_{\seR}^2 +\tilde{l}_{R1},\quad M_{\smuR}^2 +\tilde{l}_{R2},\quad M_{\stauR}^2 +\tilde{l}_{R3}\big)
\end{eqnarray}
\end{widetext}
with $X_t=A-\mu \tan\beta$, $\tilde{l}_{Li} = Q_{\chi}^i M_{Z^{\prime}}^2c_{2\gamma} + (\frac{1}{2} -c_W^2)M_Z^2c_{2\beta}$ and $\tilde{l}_{Ri} = -Q_{\chi}^i M_{Z^{\prime}}^2c_{2\gamma} - (1 -c_W^2)M_Z^2c_{2\beta}$.

This matrix has one exactly zero eigenvalue, as expected, that gives mass to the $W^{\pm}$ gauge bosons.

\section{Chargino Mass Matrix}\label{chargino}

The mass terms in the Lagrangian corresponding to charged fermions can be written as
\begin{equation}
\mathcal{L}^{\pm}\supset -\frac{1}{2}(\psi^{-T}X\psi^+ + \psi^{+T}X^T\psi^-)+H.c.
\end{equation}
where
\begin{eqnarray}
\psi^-=(i\wmino, \hdmino, e_L^-, \mu_L^-, \tau_L^-)\nn
\psi^+=(i\wpino, \hupino, e_R^+, \mu_R^+, \tau_R^+)\nonumber
\end{eqnarray}
and $X$ is the chargino mass matrix, 
\begin{equation}
X=
\left(
\begin{array}{ccccc}
M_2&\sqrt{2}M_W s_{\beta}&0&0&0\\
\sqrt{2}M_Wc_{\beta}&\mu&0&0&0\\
0&\mu_1&m_e&0&0\\
0&\mu_2&0&m_{\mu}&0\\
0&\mu_3&0&0&m_{\tau}
\end{array}
\right)
\end{equation}

This may also be written in a more compact form by introducing the vector,
\begin{equation}
\psi_{\pm}=(\psi^-,\psi^+)
\end{equation}
and 
\begin{equation}
M_{\pm}=
\left(
\begin{array}{cc}
0&X\\
X^T&0
\end{array}
\right)
\end{equation}
such that,
\begin{equation}
\mathcal{L}^{\pm}=-\frac{1}{2}\psi_{\pm}^TM_{\pm}\psi_{\pm} + h.c.
\end{equation}

In order to diagonalize the mass matrix $X$ we need two matrices--one that transforms $\psi^-$ ($U$) and another that transforms $\psi^+$ ($V$)--so that
\begin{equation}
U^{\ast}XV^{\dag}=m_{{\tilde{\chi}}^{\pm}}.
\end{equation} 
The matrices $U$ and $V$ diagonalize the matrices $X X^T$ and $X^T X$ respectively. The charged leptons e, $\mu$ and $\tau$ also enter our chargino mass matrix and, in general, mix with the wino and Higgsino. However, this mixing is extremely weak and hence the relevant mixing matrix elements in the calculations may be taken to be unity. It is for this reason that they also do not enter into our neutrino mass matrix calculation.

\section{Charged Lepton Flavor Violation}\label{lfvapp}

We follow and extend the calculation for $l_j\rightarrow l_i\gamma$ amplitudes and couplings given in Ref.\cite{Carvalho:2002bq}. The couplings for the calculation are 
defined below. We denote the sign of the neutralino mass eigenvalue by $\epsilon_A$.

\subsection{Chargino-chargino-neutral scalar couplings}

\begin{eqnarray}
V^{ccs}_{LiAX}&=& \frac{1}{\sqrt{2}}\big[V_{i4}\big(-\lambda_{122}U_{A4}U^{\ast}_{\tilde{S}^0X1} + y_{\mu} U_{A2}U^{\ast}_{\tilde{S}^0X2} \nn 
&+& \lambda_{122}U_{A3}U^{\ast}_{\tilde{S}^0X2}\big) + y_e V_{i3} U_{A2}U^{\ast}_{\tilde{S}^0X1} \nn
&+& V_{i5}\big(-\lambda_{133}U_{A5}U^{\ast}_{\tilde{S}^0X1} + y_{\tau} U_{A2}U^{\ast}_{\tilde{S}^0X3}\nn
&+& \lambda_{133}U_{A3}U^{\ast}_{\tilde{S}^0X3}\big)\big]\\
V^{ccs}_{RiAX}&=&\frac{1}{\sqrt{2}}\big[U^{\ast}_{i4}\big(-g_2 V^{\ast}_{A1}U^{\ast}_{\tilde{S}^0X2} - \lambda_{122} V^{\ast}_{A4} U^{\ast}_{\tilde{S}^0X1} \nn
&-& y_{\eta} V^{\ast}_{A2} U^{\ast}_{\tilde{S}^0X6}\big) + U^{\ast}_{i3}\big(-g_2V^{\ast}_{A1} U^{\ast}_{\tilde{S}^0X1}\nn
&+& \lambda_{122} V^{\ast}_{A4} U^{\ast}_{\tilde{S}^0X2} + \lambda_{133} V^{\ast}_{A5} U^{\ast}_{\tilde{S}^0X3}\big)\nn
&+& U^{\ast}_{i5}\big(-g_2 V^{\ast}_{A1} U^{\ast}_{\tilde{S}^0X3} -\lambda_{133}V^{\ast}_{A5} U^{\ast}_{\tilde{S}^0X1}\nn
&-& y_{\etabar}V^{\ast}_{A2} U^{\ast}_{\tilde{S}^0X7}\big)\big]
\end{eqnarray}

\subsection{Chargino-chargino-neutral pseudoscalar couplings}

\begin{eqnarray}
V^{ccp}_{LiAX}&=& \frac{i}{\sqrt{2}}\big[V_{i4}\big(-\lambda_{122}U_{A4}U^{\ast}_{\tilde{P}^0X1} + y_{\mu} U_{A2}U^{\ast}_{\tilde{P}^0X2} \nn 
&+& \lambda_{122}U_{A3}U^{\ast}_{\tilde{P}^0X2}\big) + y_e V_{i3} U_{A2}U^{\ast}_{\tilde{P}^0X1} \nn
&+& V_{i5}\big(-\lambda_{133}U_{A5}U^{\ast}_{\tilde{P}^0X1} + y_{\tau} U_{A2}U^{\ast}_{\tilde{P}^0X3}\nn
&+& \lambda_{133}U_{A3}U^{\ast}_{\tilde{P}^0X3}\big)\big]\\
V^{ccp}_{RiAX}&=&\frac{i}{\sqrt{2}}\big[U^{\ast}_{i4}\big(-g_2 V^{\ast}_{A1}U^{\ast}_{\tilde{P}^0X2} + \lambda_{122} V^{\ast}_{A4} U^{\ast}_{\tilde{P}^0X1} \nn
&+& y_{\eta} V^{\ast}_{A2} U^{\ast}_{\tilde{P}^0X6}\big) + U^{\ast}_{i3}\big(-g_2V^{\ast}_{A1} U^{\ast}_{\tilde{P}^0X1}\nn
&-& \lambda_{122} V^{\ast}_{A4} U^{\ast}_{\tilde{P}^0X2} - \lambda_{133} V^{\ast}_{A5} U^{\ast}_{\tilde{P}^0X3}\big)\nn
&+& U^{\ast}_{i5}\big(-g_2 V^{\ast}_{A1} U^{\ast}_{\tilde{P}^0X3} +\lambda_{133}V^{\ast}_{A5} U^{\ast}_{\tilde{P}^0X1}\nn
&+& y_{\etabar}V^{\ast}_{A2} U^{\ast}_{\tilde{P}^0X7}\big)\big]
\end{eqnarray}

\subsection{Chargino-neutralino-charged scalar couplings}

\begin{eqnarray}
V^{cns}_{LiAX}&=& \epsilon_A\big[V^{\ast}_{i3}\big(-\sqrt{2}g_1 N^{\ast}_{A5}U^{\ast}_{\tilde{S}^{\pm}X6} - y_e N^{\ast}_{A7}U^{\ast}_{\tilde{S}^{\pm}X3}\nn
&+& y_e N^{\ast}_{A1}U^{\ast}_{\tilde{S}^{\pm}X2}\big) + V^{\ast}_{i4}\big(-\sqrt{2}g_1 N^{\ast}_{A5}U^{\ast}_{\tilde{S}^{\pm}X7}\nn
&+& \frac{g_X}{\sqrt{2}}N^{\ast}_{A4}U^{\ast}_{\tilde{S}^{\pm}X7} - y_{\mu} N^{\ast}_{A7}U^{\ast}_{\tilde{S}^{\pm}X4}\nn
&+& y_{\mu} N^{\ast}_{A2}U^{\ast}_{\tilde{S}^{\pm}X2} -\lambda_{122} N^{\ast}_{A1}U^{\ast}_{\tilde{S}^{\pm}X4}\nn
&+& \lambda_{122} N^{\ast}_{A2}U^{\ast}_{\tilde{S}^{\pm}X3}\big) + V^{\ast}_{i5}\big(-\sqrt{2}g_1 N^{\ast}_{A5}U^{\ast}_{\tilde{S}^{\pm}X8}\nn
&-& \frac{g_X}{\sqrt{2}}N^{\ast}_{A4}U^{\ast}_{\tilde{S}^{\pm}X8} - y_{\tau} N^{\ast}_{A7}U^{\ast}_{\tilde{S}^{\pm}X5}\nn
&+& y_{\tau} N^{\ast}_{A3}U^{\ast}_{\tilde{S}^{\pm}X2} -\lambda_{133} N^{\ast}_{A1}U^{\ast}_{\tilde{S}^{\pm}X5}\nn
&+& \lambda_{133} N^{\ast}_{A3}U^{\ast}_{\tilde{S}^{\pm}X3}\big)\big]
\end{eqnarray}
\begin{eqnarray}
V^{cns}_{RiAX}&=& U^{\ast}_{i3}\big(\frac{g_1}{\sqrt{2}}N^{\ast}_{A5}U^{\ast}_{\tilde{S}^{\pm}X3} + \frac{g_2}{\sqrt{2}} N^{\ast}_{A6}U^{\ast}_{\tilde{S}^{\pm}X3}\nn
&-& y_e N^{\ast}_{A7}U^{\ast}_{\tilde{S}^{\pm}X6} + \lambda_{122} N^{\ast}_{A2}U^{\ast}_{\tilde{S}^{\pm}X7}\nn
&+& \lambda_{133} N^{\ast}_{A3}U^{\ast}_{\tilde{S}^{\pm}X8}\big) + U^{\ast}_{i4} \big(\frac{g_1}{\sqrt{2}} N^{\ast}_{A5}U^{\ast}_{\tilde{S}^{\pm}X4}\nn
&+& \frac{g_2}{\sqrt{2}}N^{\ast}_{A6}U^{\ast}_{\tilde{S}^{\pm}X4} - \frac{g_X}{\sqrt{2}}N^{\ast}_{A4}U^{\ast}_{\tilde{S}^{\pm}X4}\nn
&-& y_{\eta} N^{\ast}_{A9}U^{\ast}_{\tilde{S}^{\pm}X1} - y_{\mu} N^{\ast}_{A7}U^{\ast}_{\tilde{S}^{\pm}X7}\nn
&-& \lambda_{122} N^{\ast}_{A1}U^{\ast}_{\tilde{S}^{\pm}X7}\big) + U^{\ast}_{i5}\big(\frac{g_1}{\sqrt{2}} N^{\ast}_{A5}U^{\ast}_{\tilde{S}^{\pm}X5}\nn
&+& \frac{g_2}{\sqrt{2}}N^{\ast}_{A6}U^{\ast}_{\tilde{S}^{\pm}X5} + \frac{g_X}{\sqrt{2}}N^{\ast}_{A4}U^{\ast}_{\tilde{S}^{\pm}X5}\nn
&-& y_{\etabar} N^{\ast}_{A10}U^{\ast}_{\tilde{S}^{\pm}X1} - y_{\tau} N^{\ast}_{A7}U^{\ast}_{\tilde{S}^{\pm}X8}\nn
&-& \lambda_{133} N^{\ast}_{A1}U^{\ast}_{\tilde{S}^{\pm}X8}\big)
\end{eqnarray}

\subsection{Chargino-chargino-$Z$ couplings}

\begin{eqnarray}
V^{ccZ}_{LiA} &=& \frac{g_2}{c_W} \big(\frac12 U_{i1}U^{\ast}_{A1} + \left(\frac12 - s^2_W\right)\delta_{iA}\big)\\
V^{ccZ}_{RiA} &=& \frac{g_2}{c_W} \big(V^{\ast}_{i1}V_{A1} + \frac12 V^{\ast}_{i2} V_{A2} -s^2_W \delta_{iA}\big)
\end{eqnarray}

\subsection{Chargino-chargino-$Z^{\prime}$ couplings}

\begin{eqnarray}
V^{ccZ^{\prime}}_{LiA} &=& \frac{g_X}{2}\big(U_{i5}U^{\ast}_{A5}-U_{i4}U^{\ast}_{A4}\big)\\
V^{ccZ^{\prime}}_{RiA} &=& \frac{g_X}{2}\big(V^{\ast}_{i5}V_{A5}-V^{\ast}_{i4}V_{A4}\big)
\end{eqnarray}

\subsection{Chargino-neutralino-$W$ couplings}

\begin{eqnarray}
V^{cnW}_{LiA} &=& -\epsilon_A g_2 \big[U_{i1}N^{\ast}_{A6} + \frac{1}{\sqrt{2}}\big(U_{i2}N^{\ast}_{A7}\nn
&+& U_{i3}N^{\ast}_{A1} + U_{i4}N^{\ast}_{A2} + U_{i5}N^{\ast}_{A3}\big)\big]\\
V^{cnW}_{RiA} &=& g_2\big[\frac{1}{\sqrt{2}} V^{\ast}_{i2} N_{A8} -V^{\ast}_{i1} N_{A6}\big]
\end{eqnarray}

\subsection{The neutralino-charged scalar amplitude}
We provide the expression for the left-handed amplitude only; the corresponding right-handed amplitude is given by $A^R_{ij}$ = $A^L_{ij}$ (L/R$\rightarrow$R/L).
 
\begin{eqnarray}
A^{L,\tilde{\chi}^0\tilde{S}^{\pm}}_{ij} &=& \sum_{A=1}^{10}\sum_{X=1}^{7} \frac{1}{32\pi^2}\frac{1}{m^2_{\tilde{S}^{\pm}_X}}\nn
&&\bigg[f_n\left(x^n_{AX}\right)V^{cns}_{LiAX}V^{cns\ast}_{LjAX}\nn
&+& h_n\left(x^n_{AX}\right)\frac{m_{\tilde{\chi}^0_A}}{m_{l_j}}V^{cns}_{LiAX}V^{cns\ast}_{RjAX}\bigg],
\end{eqnarray}
where
\begin{equation}
x^n_{AX}=\left(\frac{m_{\tilde{\chi}^0_A}}{m_{\tilde{S}^{\pm}_X}}\right)^2
\end{equation}
and 
\begin{eqnarray}
f_n(x)&=& \frac{1-6x+3x^2+2x^3-6x^2\log x}{6(1-x)^4},\nn
h_n(x)&=& \frac{1-x^2+2x\log x}{(1-x)^3}.
\end{eqnarray}

\subsection{The chargino-neutral scalar amplitude}

The amplitude is
\begin{eqnarray}
A^{L,\tilde{\chi}^{\pm}\tilde{S}^0}_{ij} &=& \sum_{A=1}^{5}\sum_{X=1}^{7} - \frac{1}{32\pi^2}\frac{1}{m^2_{\tilde{S}^{0}_X}}\nn
&&\bigg[f_c\left(x^{cs}_{AX}\right)V^{ccs}_{LiAX}V^{ccs\ast}_{LjAX}\nn
&+& h_c\left(x^{cs}_{AX}\right)\frac{m_{\tilde{\chi}^{\pm}_A}}{m_{l_j}}V^{ccs}_{LiAX}V^{ccs\ast}_{RjAX}\bigg].
\end{eqnarray}
We use the definitions
\begin{eqnarray}
x^{cs}_{AX}&=&\left(\frac{m_{\tilde{\chi}^{\pm}_A}}{m_{\tilde{S}^{0}_X}}\right)^2 ,\nn
f_c(x)&=& \frac{2+3x-6x^2+x^3+6x\log x}{6(1-x)^4},\nn
h_c(x)&=& \frac{-3+4x-x^2-2\log x}{(1-x)^3}.
\end{eqnarray}

\subsection{The chargino-neutral pseudoscalar amplitude}
The left-handed amplitude for the chargino-neutral pseudoscalar loop is given by
\begin{eqnarray}
A^{L,\tilde{\chi}^{\pm}\tilde{P}^0}_{ij} &=& \sum_{A=1}^{5}\sum_{X=1}^{5} - \frac{1}{32\pi^2}\frac{1}{m^2_{\tilde{P}^{0}_X}}\nn
&&\bigg[f_c\left(x^{cp}_{AX}\right)V^{ccp}_{LiAX}V^{ccp\ast}_{LjAX}\nn
&+& h_c\left(x^{cp}_{AX}\right)\frac{m_{\tilde{\chi}^{\pm}_A}}{m_{l_j}}V^{ccp}_{LiAX}V^{ccp\ast}_{RjAX}\bigg],
\end{eqnarray}
where,
\begin{eqnarray}
x^{cp}_{AX}&=&\left(\frac{m_{\tilde{\chi}^{\pm}_A}}{m_{\tilde{P}^{0}_X}}\right)^2.
\end{eqnarray}

\subsection{The chargino-$Z^0$ boson amplitude}

The amplitude is defined to be
\begin{eqnarray}
A^{L,Z\tilde{\chi}^{\pm}}_{ij} &=& \sum_{A=1}^{5} \frac{1}{32\pi^2}\frac{1}{M^2_{Z}}\nn
&&\bigg[f_z\left(x^z_{A}\right)V^{ccZ}_{RiA}V^{ccZ\ast}_{RjA}\nn
&+& h_z\left(x^z_{A}\right)\frac{m_{\tilde{\chi}^{\pm}_A}}{m_{l_j}}V^{ccZ}_{RiA}V^{ccZ\ast}_{LjA}\bigg],
\end{eqnarray}
with
\begin{eqnarray}
x^z_A&=&\left(\frac{m_{\tilde{\chi}^{\pm}_A}}{M_Z}\right)^2 ,\nn 
f_z(x)&=& \frac{8-38x+39x^2-14x^3+5x^4-18x^2\log x}{6(1-x)^4},\nn
h_z(x)&=& \frac{-4+3x+x^3-6x\log x}{(1-x)^3}.
\end{eqnarray}

\subsection{The chargino-$Z^{\prime}$ boson amplitude}

This amplitude is given by
\begin{eqnarray}
A^{L,Z^{\prime}\tilde{\chi}^{\pm}}_{ij} &=& \sum_{A=1}^{5} \frac{1}{32\pi^2}\frac{1}{M^2_{Z^{\prime}}}\nn
&&\bigg[f_z\left(x^{z^{\prime}}_{A}\right)V^{ccZ^{\prime}}_{RiA}V^{ccZ^{\prime}\ast}_{RjA}\nn
&+& h_z\left(x^{z^{\prime}}_{A}\right)\frac{m_{\tilde{\chi}^{\pm}_A}}{m_{l_j}}V^{ccZ^{\prime}}_{RiA}V^{ccZ^{\prime}\ast}_{LjA}\bigg]
\end{eqnarray}
with
\begin{eqnarray}
x^{z^{\prime}}_A &=& \left(\frac{m_{\tilde{\chi}^{\pm}_A}}{M_{Z^{\prime}}}\right)^2.
\end{eqnarray}

\subsection{The neutralino-$W^{\pm}$ boson amplitude}
The amplitude is
\begin{eqnarray}
A^{L,W\tilde{\chi}^{0}}_{ij} &=& \sum_{A=1}^{10} - \frac{1}{32\pi^2}\frac{1}{M^2_{W}}\nn
&&\bigg[f_w\left(x^w_{A}\right)V^{cnW}_{RiA}V^{cnW\ast}_{RjA}\nn
&+& h_w\left(x^w_{A}\right)\frac{m_{\tilde{\chi}^{0}_A}}{m_{l_j}}V^{cnW}_{RiA}V^{cnW\ast}_{LjA}\bigg],
\end{eqnarray}
with
\begin{eqnarray}
x^w_A&=& \left(\frac{m_{\tilde{\chi}^{0}_A}}{M_W}\right)^2,\nn
f_w(x)&=& \frac{10-43x+78x^2-49x^3+4x^4+18x^3\log x}{6(1-x)^4},\nn
h_w(x)&=& \frac{-4+15x-12x^2+x^3+6x^2\log x}{(1-x)^3}.
\end{eqnarray}

\end{document}